\documentclass[preprint,10pt]{sigplanconf}

\usepackage{amsthm}
\usepackage{amsmath}
\usepackage{amssymb}
\usepackage{graphicx}
\usepackage{epstopdf}
\usepackage{hyperref}
\usepackage{alltt}
\usepackage{listings}
\usepackage{array}
\usepackage{extarrows}
\usepackage{setspace}
\usepackage{tikz}
\usepackage{tikz-qtree}
\usetikzlibrary{calc}
\usetikzlibrary{positioning}
\usepackage{floatrow}
\usepackage{mathtools}
\usepackage{makecell}
\usepackage[noline, boxed, procnumbered, linesnumberedhidden, titlenumbered]{algorithm2e} 
\usepackage{enumitem}

\newcommand{\techreport}[1]{} 

\newcommand{\ith}{$i^{th}$}
\newcommand{\cut}[1]{}
\newcommand{\todo}[1]{{\bf\em TODO:} #1}
\newcommand*{\pos}[1]{\ensuremath{\color{blue}\uparrow_{\!#1}}}
\newcommand*{\chpos}[2]{\ensuremath{\underset{\color{blue}\uparrow_{\!#2}}{#1}}}

\cut{

}

\newcommand{\codebuff}{\textsc{\small CodeBuff}\xspace}

\newcommand\vizsp[1][.33em]{%
  \mbox{\kern.06em\vrule height.3ex}%
  \vbox{\hrule width#1}%
  \hbox{\vrule height.3ex}}
    
\newtheorem{definition}{Definition}[section]
\newcounter{featurecounter}

\setlist[enumerate]{itemsep=-1mm}

\cut{
\addtolength{\topmargin}{-17pt}
\addtolength{\textfloatsep}{-10pt}
\addtolength{\textheight}{12pt}
\addtolength{\intextsep}{-3pt}
\addtolength{\oddsidemargin}{-4pt}
\addtolength{\evensidemargin}{-4pt}

\addtolength{\topsep}{-4pt}
}

\begin{document}

\setlength{\pdfpageheight}{\paperheight}
\setlength{\pdfpagewidth}{\paperwidth}

\conferenceinfo{SLE '16}{October 31--November 1, 2016, City, ST, Country}
\copyrightyear{2016}
\copyrightdata{978-1-nnnn-nnnn-n/yy/mm}


\title{Technical Report: Towards a Universal Code Formatter through Machine Learning}

\authorinfo{Terence Parr\vspace{-2mm}}
           {University of San Francisco\vspace{-1mm}}
           {parrt@cs.usfca.edu}
\authorinfo{Jurgen Vinju\vspace{-2mm}}
           {Centrum Wiskunde \& Informatica\vspace{-1mm}}
           {Jurgen.Vinju@cwi.nl}

\maketitle

\begin{abstract}
There are many declarative frameworks that allow us to implement code formatters relatively easily for any specific language, but constructing them is cumbersome. The first problem is that ``everybody'' wants to format their code differently, leading to either many formatter variants or a ridiculous number of configuration options. Second, the size of each  implementation scales with a language's grammar size, leading to hundreds of rules. 

In this paper, we solve the formatter construction problem using a novel approach, one that automatically derives formatters for any given language without intervention from a language expert. We introduce a code formatter called \codebuff that uses machine learning to abstract formatting rules from a representative corpus, using a carefully designed feature set.  Our experiments on Java, SQL, and ANTLR grammars show that \codebuff is efficient, has excellent accuracy, and is grammar invariant for a given language. It also generalizes to a 4th language tested during manuscript preparation.
\end{abstract}



\section{Introduction\label{sec:intro}\label{sec:introduction}}

The way source code is formatted has a significant impact on its comprehensibility~\cite{MMNS83}, and manually reformatting code is just not an option~\cite[p.399]{mcconnel93}. Therefore, programmers need ready access to automatic code formatters or ``pretty printers'' in situations where formatting is messy or inconsistent. Many program generators also take advantage of code formatters to improve the quality of their output.

Because the value of a particular code formatting style is a subjective notion, often leading to heated discussions, formatters must be highly configurable.  This allows, for example, current maintainers of existing code to improve their effectiveness by reformatting the code per their preferred style. There are plenty of configurable formatters for existing languages, whether in IDEs like Eclipse or standalone tools like Gnu {\tt indent}, but specifying style is not easy. The emergent behavior is not always obvious, there exists interdependency between options, and the tools cannot take context information into account \cite{kooiker}. For example, here are the options needed to obtain K\&R C style with {\tt indent}:

\begin{small}
\begin{verbatim}
-nbad -bap -bbo -nbc -br -brs -c33 -cd33 -ncdb -ce
-ci4 -cli0 -cp33 -cs -d0 -di1 -nfc1 -nfca -hnl -i4 
-ip0 -l75 -lp -npcs -nprs -npsl -saf -sai -saw -nsc
-nsob -nss
\end{verbatim}
\end{small}

New languages pop into existence all the time and each one could use a formatter. Unfortunately, building a formatter is difficult and tedious. Most formatters used in practice are {\em ad hoc}, language-specific programs but there are formal approaches that yield good results with less effort. Rule-based formatting systems let programmers specify phrase-formatting pairs, such as the following sample specification for formatting the COBOL {\tt MOVE} statement using ASF+SDF~\cite{formatter-gen,metaenv,box,kooiker}.

\begin{small}
\begin{verbatim}
MOVE IdOrLit TO Id-list =
from-box( H [ "MOVE"
              H ts=25 [to-box(IdOrLit)]
              H ts=49 ["TO"]
              H ts=53 [to-box(Id-list)] ])
\end{verbatim}
\end{small}
\noindent This rule maps a parse tree pattern to a box expression. A set of such rules, complemented with default behavior for the unspecified parts, generates a single formatter with a specific style for the given language. Section \ref{sec:related} has other related work.

There are a number of problems with rule-based formatters. First, each specification yields a formatter for one specific style.  Each new style requires a change to those rules or the creation of a new set. Some systems allow the rules to be parametrized, and configured accordingly, but that leads to higher rule complexity. Second, minimal changes to the associated grammar usually require changes to the formatting rules, even if the grammar changes do not affect the language recognized. Finally, formatter specifications are big. Although most specification systems have builtin heuristics for default behavior in the absence of a specification for a given language phrase, specification size tends to grow with the grammar size. A few hundred rules are no exception.

Formatting is a problem solved in theory, but not yet in practice.  Building a good code formatter is still too difficult and requires way too much work; we need a fresh approach.
In this paper, we introduce a tool called \codebuff~\cite{codebuff} that uses machine learning to produce a formatter entirely from a grammar for language $L$ and a representative corpus written in $L$.  There is no specification work needed from the user other than to ensure reasonable formatting consistency within the corpus. The statistical model used by \codebuff first learns the formatting rules from the corpus, which are then applied to format other documents in the same style. Different corpora effectively result in different formatters. From a user perspective the formatter is ``configured by example.''

\paragraph{Contributions and roadmap.}  We begin by showing sample \codebuff output in Section~\ref{sec:examples} and then explain how and why \codebuff works in Section~\ref{sec:design}. Section~\ref{sec:evaluation} provides empirical evidence that \codebuff learns a formatting style quickly and using very few files. \codebuff approximates the corpus style with high accuracy for the languages ANTLR, Java and SQL, and it is largely insensitive to language-preserving grammar changes. To adjust for possible selection bias and model overfitting to these three well-known languages, we tested \codebuff on an unfamiliar language (Quorum) in Section~\ref{sec:validation}, from which we learned that \codebuff works similarly well, yet improvements are still possible. We position \codebuff with respect to the literature on formatting in Section~\ref{sec:related_work}.

\section{Sample Formatting\label{sec:examples}\label{sec:motivatingExamples}}

This section contains sample SQL, Java, and ANTLR code formatted by \codebuff, including some that are poorly formatted to give a balanced presentation.  Only the  formatting style matters here so we use a small font for space reasons.  Github \cite{codebuff} has a snapshot of all input corpora and formatted versions (corpora, testing details in Section \ref{empirical}). To arrive at the formatted output for document $d$ in corpus $D$, our test rig removes all whitespace tokens from $d$ and then applies an instance of \codebuff trained on the corpus without $d$, $D \setminus \{d\}$.  

The examples are not meant to illustrate ``good style.'' They are simply consistent with the style of a specific corpus.  In Section~\ref{sec:evaluation} we define a metric to measure the  success of the automated formatter in an objective and reproducible manner.  No quantitative research method can capture the qualitative notion of style,  so we start with these examples. (We use ``\ldots'' for immaterial text removed to shorten samples.)

SQL is notoriously difficult to format, particularly for nested queries, but \codebuff does an excellent job in most cases. For example, here is a formatted query from file {\small IPMonVerificationMaster.sql} (trained with {\tt sqlite} grammar on {\tt sqlclean} corpus):

\begin{tiny}
\begin{verbatim}
SELECT DISTINCT
    t.server_name
       , t.server_id
       , 'Message Queuing Service' AS missingmonitors
FROM t_server t INNER JOIN t_server_type_assoc tsta ON t.server_id = tsta.server_id
WHERE t.active = 1 AND tsta.type_id IN ('8')
      AND t.environment_id = 0
      AND t.server_name NOT IN
    (
          SELECT DISTINCT l.address
          FROM ipmongroups g INNER JOIN ipmongroupmembers m ON g.groupid = m.groupid
               INNER JOIN ipmonmonitors l ON m.monitorid = l.monitorid
               INNER JOIN t_server t ON l.address = t.server_name
               INNER JOIN t_server_type_assoc tsta ON t.server_id = tsta.server_id
          WHERE l.name LIKE '%Message Queuing Service%'
                AND t.environment_id = 0
                AND tsta.type_id IN ('8')
                AND g.groupname IN ('Prod O/S Services')
                AND t.active = 1
      )
UNION
ALL
\end{verbatim}
\end{tiny}

\noindent And here is a complicated query from 
{\scriptsize dmart\_bits\_IAPPBO510.sql} with case statements:

\begin{tiny}
\begin{verbatim}
SELECT
    CASE WHEN SSISInstanceID IS NULL
        THEN 'Total'
    ELSE SSISInstanceID END SSISInstanceID
    , SUM(OldStatus4) AS OldStatus4
    ...
    , SUM(OldStatus4 + Status0 + Status1 + Status2 + Status3 + Status4) AS InstanceTotal
FROM
    (
        SELECT
            CONVERT(VARCHAR, SSISInstanceID)             AS SSISInstanceID
            , COUNT(CASE WHEN Status = 4 AND
                              CONVERT(DATE, LoadReportDBEndDate) <
                              CONVERT(DATE, GETDATE())
                        THEN Status
                    ELSE NULL END)             AS OldStatus4
            ...
            , COUNT(CASE WHEN Status = 4 AND
                              DATEPART(DAY, LoadReportDBEndDate) = DATEPART(DAY, GETDATE())
                        THEN Status
                    ELSE NULL END)             AS Status4
        FROM dbo.ClientConnection
        GROUP BY SSISInstanceID
    ) AS StatusMatrix
GROUP BY SSISInstanceID\end{verbatim}
\end{tiny}

\noindent Here is a snippet from Java, our 2nd test language, taken from {\small STLexer.java} (trained with {\tt java} grammar on {\tt st} corpus):

\begin{tiny}
\begin{verbatim}
switch ( c ) {
    ...
    default:
        if ( c==delimiterStopChar ) {
            consume();
            scanningInsideExpr = false;
            return newToken(RDELIM);
        }
        if ( isIDStartLetter(c) ) {
            ...
            if ( name.equals("if") ) return newToken(IF);
            else if ( name.equals("endif") ) return newToken(ENDIF);
            ...
            return id;
        }
        RecognitionException re = new NoViableAltException("", 0, 0, input);
        ...
        errMgr.lexerError(input.getSourceName(),
                          "invalid character '"+str(c)+"'",
                          templateToken,
                          re);
        ...
\end{verbatim}
\end{tiny}

Here is an example from {\tt STViz.java} that indents a method declaration relative to the start of an expression rather than the first token on the previous line:

\begin{tiny}
\begin{verbatim}
Thread t = new Thread() {
               @Override
               public void run() {
                   synchronized ( lock ) { 
                       while ( viewFrame.isVisible() ) { 
                           try {
                               lock.wait();
                           }   
                           catch (InterruptedException e) {
                           }   
                       }   
                   }   
               }   
           };  
\end{verbatim}
\end{tiny}

\noindent Formatting results are generally excellent for ANTLR, our third test language. E.g., here is a snippet from {\small Java.g4}:

\begin{tiny}
\begin{verbatim}
classOrInterfaceModifier
    :   annotation       // class or interface
    |   (   'public'     // class or interface
            ...
        |   'final'      // class only -- does not apply to interfaces
        |   'strictfp'   // class or interface
        )
    ;
\end{verbatim}
\end{tiny}

\noindent Among the formatted files for the three languages, there are a few regions of inoptimal or bad formatting. \codebuff does not capture all formatting rules and occasionally gives puzzling formatting. For example, in the {\small Java8.g4} grammar, the following rule has all elements packed onto one line (``$\hookleftarrow$'' means we soft-wrapped output for printing purposes):

\begin{tiny}
\begin{alltt}
unannClassOrInterfaceType
    :   (unannClassType_lfno_unannClassOrInterfaceType | \(\hookleftarrow\)
unannInterfaceType_lfno_unannClassOrInterfaceType) \(\hookleftarrow\)
(unannClassType_lf_unannClassOrInterfaceType |\(\hookleftarrow\)
unannInterfaceType_lf_unannClassOrInterfaceType)*
    ;
\end{alltt}
\end{tiny}

\noindent \codebuff does not consider line length during training or formatting, instead mimicking the natural line breaks found among phrases of the corpus. For Java and SQL this works very well, but not always with ANTLR grammars.

Here is an interesting Java formatting issue from {\small Compiler.java}  that is indented too far to the right (column 102); it is indented from the {\tt \{\{}.  That is a good decision in general, but here the left-hand side of the assignment is very long, which indents the {\tt put()} code too far to be considered good style.

\begin{tiny}
\begin{verbatim}
public ... Map<...> defaultOptionValues = new HashMap<...>() {{
                                                                  put("anchor", "true");
                                                                  put("wrap", "\n");
                                                              }};
\end{verbatim}
\end{tiny}

\noindent In {\small STGroupDir.java}, the {\tt prefix} token is aligned improperly:

\begin{tiny}
\begin{verbatim}
if ( verbose ) System.out.println("loadTemplateFile("+unqualifiedFileName+") in groupdir...
                                  " prefix=" +
prefix);
\end{verbatim}
\end{tiny}

We also note that some of the SQL expressions are incorrectly aligned, as in this sample from {\small SQLQuery23.sql}:
 
\begin{tiny}
\begin{verbatim}
AND dmcl.ErrorMessage NOT LIKE '%Pre-Execute phase is beginning.  %'
                               AND dmcl.ErrorMessage NOT LIKE '%Prepare for Execute phase...
                                                              AND dmcl.ErrorMessage NOT...
\end{verbatim}
\end{tiny}

Despite a few anomalies, \codebuff generally reproduces a corpus' style well. Now we describe the design used to achieve these results. In Section~\ref{sec:evaluation} we quantify them.

\newcommand{\WS}{\ensuremath{\mathit{ws}}\xspace}
\newcommand{\NL}{\ensuremath{\mathit{nl}}\xspace}
\newcommand{\SPACE}{\ensuremath{\mathit{sp}}\xspace}
\newcommand{\ALIGN}{\ensuremath{\mathit{align}}\xspace}
\newcommand{\INDENT}{\ensuremath{\mathit{indent}}\xspace}
\newcommand{\HPOS}{\ensuremath{\mathit{hpos}}\xspace}
\newcommand{\NONE}{\ensuremath{\mathit{none}}\xspace}
\newcommand{\TOKEN}{\ensuremath{\mathit{tok}}\xspace}

\section{The Design of an AI for Formatting\label{sec:design}}

Our AI formatter mimics what programmers do during the act of entering code.  Before entering a program symbol, a programmer decides {\em (i)} whether a space or line break is required and if line break, {\em (ii)} how far to indent the next line. 
Previous approaches (see Section~\ref{sec:related_work}) make a language engineer define whitespace injection programmatically.

A formatting engine based upon machine learning operates in two distinct phases: {\em training} and {\em formatting}. The {\em training} phase examines a corpus of code documents, $D$, written in language $L$ to construct a statistical {\em model} that represents the formatting style of the corpus author. The essence of training is to capture the whitespace preceding each token, $t$, and then associate that whitespace with the phrase context surrounding $t$.  Together, the context and whitespace preceding $t$ form an {\em exemplar}.  Intuitively, an exemplar captures how the corpus author formatted a specific, fine-grained piece of a phrase, such as whether the author placed a newline before or after the left curly brace in the context of a Java {\tt if}-statement.  

Training captures the context surrounding $t$ as an $m$-dimensional {\em feature vector}, $X$, that includes $t$'s token type, parse-tree ancestors, and many other {\em features} (Section \ref{features}). Training captures the whitespace preceding $t$ as the concatenation of two separate  operations or {\em directives}: a whitespace \WS  directive followed by a horizontal positioning \HPOS  directive if \WS is a newline (line break). The \WS  directive generates spaces, newlines, or nothing while \HPOS generates spaces to indent or align $t$ relative to a previous token (Section \ref{sec:directives}).  

As a final step, training presents the list of exemplars to a machine learning algorithm that constructs a statistical model.  There are $N$ exemplars $(X_j,w_j,h_j)$ for $j=1..N$ where $N$ is the number of total tokens in all documents of corpus $D$ and $w_j \in \WS$, $h_j \in \HPOS$.  Machine learning models are typically both a highly-processed condensation of the exemplars and a {\em classifier function} that, in our case, classifies a context feature vector, $X$, as needing a specific bit of whitespace. Classifier functions predict how the corpus author would format a specific context by returning a formatting directive.  A model for formatting needs two classifier functions, one for predicting \WS and one for \HPOS (consulted if \WS prediction yields a newline).

\codebuff uses a $k$-Nearest Neighbor ({\em kNN}) machine learning model, which conveniently uses the list of exemplars as the actual model. A {\em kNN}'s classifier function compares an unknown context vector $X$ to the $X_j$ from all $N$ exemplars and finds the $k$ nearest.  Among these $k$, the classifier predicts the formatting directive that appears most often (details in Section \ref{sec:prediction}).  It's akin to asking a programmer how they normally format the code in a specific  situation. Training requires a corpus $D$ written in $L$, a lexer and parser for $L$ derived from grammar $G$, and the corpus indentation size to identify indented phrases; e.g., one of the Java corpora we tested indents with 2 not 4 spaces.  Let $F_{D,G}=({\bf X}, W, H, \operatorname{indentSize})$ denote the formatting model contents with context vectors forming rows of matrix ${\bf X}$ and formatting directives forming elements of vectors $W$ and $H$. Function \ref{train} embodies the training process, constructing $F_{D,G}$.

\setlength{\algomargin}{3pt}
\SetAlCapSkip{-10pt}
\begin{function}[]
\SetAlgorithmName{Alg.}{List of Algorithms}
\SetAlgoSkip{}
\SetInd{.5em}{.5em}
\TitleOfAlgo{$train$($D,G,\operatorname{indentSize}$) $\rightarrow$ model $F_{D,G}$ \vspace{-3pt}}
{\bf X} := []; $W$ := []; $H$ := []; $j$ := 1\;
\ForEach{document $d \in D$}{
{\em tokens} := tokenize($d$)\;
{\em tree} := parse({\em tokens})\;
\ForEach{$t_i$ $\in$ tokens}{
{\bf X}$[j]$ := compute context feature vector for $t_i$, {\em tree}\;
$W[j]$ := {\em capture\_ws}($t_i$)\;
$H[j]$ := {\em capture\_hpos}($t_i,\operatorname{indentSize}$)\;
$j$ := $j+1$\;
}
}
\Return{({\bf X}, $W, H, \operatorname{indentSize}$)}\;
\vspace{-2pt}
\label{train}
\end{function}

Once the model is complete, the {\em formatting} phase can begin. Formatting operates on a single document $d$ to be formatted and functions with guidance from the model. At each token $t_i \in d$, formatting computes the feature vector $X_i$ representing the context surrounding $t_i$, just like training does, but does not add $X_i$ to the model. Instead, the formatter presents $X_i$ to the \WS classifier and asks it to predict a \WS directive for $t_i$ based upon how similar contexts were formatted in the corpus.  The formatter ``executes'' the directive and, if a newline, presents $X_i$ to the \HPOS classifier to get an indentation or alignment directive. After emitting any preceding whitespace, the formatter emits the text for $t_i$. Note that any token $t_i$ is identified by its token type, string content, and offset within a specific document, $i$.

The greedy, ``local'' decisions made by the formatter give ``globally'' correct formatting results; selecting features for the $X$ vectors is critical to this success. Unlike typical machine learning tasks, our predictor functions do not yield trivial categories like ``it's a cat.'' Instead, the predicted \WS and \HPOS directives are parametrized.  The following sections detail how \codebuff captures whitespace, computes feature vectors, predicts directives, and formats documents.

\subsection{Capturing whitespace as directives}
\label{sec:directives}

In order to reproduce a particular style, formatting directives must encode the information necessary to reproduce whitespace encountered in the training corpus.  There are five canonical formatting directives:

\begin{enumerate}
\item \NL: Inject newline
\item \SPACE: Inject space character
\item $(\ALIGN, t)$: Left align current token with previous token $t$
\item $(\INDENT, t)$: Indent current token from previous token $t$
\item \NONE:  Inject nothing, no indentation, no alignment
\end{enumerate}

For simplicity and efficiency, prediction for \NL and \SPACE operations can be merged into a single ``predict whitespace'' or \WS operation and prediction of \ALIGN and \INDENT can be merged into a single ``predict horizontal position'' or \HPOS operation.  While the formatting directives are 2- and 3-tuples (details below), we pack the tuples into 32-bit integers for efficiency, $w$ for \WS directives and $h$ for \HPOS.  

\setlength{\algomargin}{3pt}
\SetAlCapSkip{-10pt}
\begin{function}[]
\SetAlgorithmName{Alg.}{List of Algorithms}
\SetAlgoSkip{}
\SetInd{.5em}{.5em}
\TitleOfAlgo{\em capture\_ws($t_i$) $\rightarrow w \in ws$ \vspace{-3pt}}
newlines := num newlines between $t_{i-1}$ and $t_i$\;
\lIf{newlines $>$ 0}{
  \Return{(\NL, newlines)}\;
}
$col\Delta$ := $t_i$.col - ($t_{i-1}$.col + len(text($t_{i-1}$)))\;
\Return{($ws, col\Delta$)}\;
\vspace{-2pt}
\label{capture-ws}
\end{function}

For $ws$ operations, the formatter needs to know how many ($n$) characters to inject: $\mathit{ws} \in \{(\mathit{nl},n), (\mathit{sp},n), \NONE\}$ as shown in Function \ref{capture-ws}. For example, in the following Java fragment, the proper $ws$ directive at \pos{a} is $(\mathit{sp},1)$, meaning ``inject 1 space,'' the directive at \pos{b} is \NONE, and \pos{c} is $(\mathit{nl},1)$, meaning ``inject 1 newline.''

\begin{alltt}\small
x\vizsp\chpos{\mathtt{=}}{a}\vizsp{}{y}\!\chpos{\mathtt{;}}{b}
\chpos{\mathtt{z}}{c}++;
\end{alltt}

The $\mathit{hpos}$ directives align or indent token $t_i$ relative to some previous token, $t_j$ for $j<i$ as computed by Function \ref{capture-hpos}. When a suitable $t_j$ is unavailable, there are $hpos$ directives that implicitly align or indent $t_i$ relative to the first token of the previous line:

$hpos \in \{ (\mathit{align,t_j}), (\mathit{indent,t_j}), \mathit{align}, \mathit{indent} \}$

\noindent In the following Java fragments, assuming 4-space indentation, directive $(indent,\mathtt{if})$ captures the whitespace at position \pos{a}, $(align,\mathtt{if})$ captures \pos{b}, and $(align,\mathtt{x})$ captures \pos{c}. 

\vspace{2mm}

\noindent\begin{tabular}{ c c c }
\begin{minipage}[t]{.3\linewidth}
\begin{alltt}\small
if\vizsp(\vizsp{}b\vizsp)\vizsp\{
\vizsp\vizsp\vizsp\vizsp{}\chpos{\mathtt{z}}{a}++;
\chpos{\mathtt{\}}}{b}
\end{alltt}
\end{minipage} &
\begin{minipage}[t]{.2\linewidth}
\begin{alltt}\small
f(x, 
\vizsp\vizsp\chpos{\mathtt{y}}{c})
\end{alltt}
\end{minipage} & 
\begin{minipage}[t]{.4\linewidth}
\begin{alltt}\small
for (int i=0; ... 
\vizsp\vizsp\vizsp\vizsp\chpos{\mathtt{x}}{d}=i;
\end{alltt}
\end{minipage}\\ 
\end{tabular}

\vspace{2mm}

\noindent At position \pos{d}, both $(\INDENT,\mathtt{for})$ and $(\ALIGN,\text{`('})$ capture the formatting, but training chooses indentation over alignment directives when both are available. We experimented with the reverse choice, but found this choice better. Here, $(\ALIGN,\text{`('})$ inadvertently captures the formatting because {\tt for}  happens to be 3 characters.

\setlength{\algomargin}{3pt}
\SetAlCapSkip{-10pt}
\begin{function}[]
\SetAlgorithmName{Alg.}{List of Algorithms}
\SetAlgoSkip{}
\SetInd{.5em}{.5em}
\TitleOfAlgo{\em capture\_hpos($t_i, \operatorname{indentSize}$) $\rightarrow h \in hpos$ \vspace{-3pt}}
ancestor := leftancestor($t_i$)\;
\uIf{$\exists$ ancestor w/child aligned with $t_i$.col}{
  $h_{align}$ := $(\ALIGN,ancestor\Delta,childindex)$\\
   \hspace{14mm}with smallest $ancestor\Delta ~\&~ childindex$\;
}
\uIf{$\exists$ ancestor w/child at $t_i$.col + $\operatorname{indentSize}$}{
  $h_{indent}$ := $(\INDENT,ancestor\Delta,childindex)$\\
   \hspace{16mm}with smallest $ancestor\Delta ~\&~ childindex$\;
}
\uIf{$h_{align}$ and $h_{indent}$ not nil}{
  \Return{directive with smallest $ancestor\Delta$}\;
}
\lIf{$h_{align}$ not nil}{\Return{$h_{align}$\;}}
\lIf{$h_{indent}$ not nil}{\Return{$h_{indent}$\;}}
\lIf{$t_i$ indented from previous line}{
  \Return{indent\;}
}
\Return{align\;}
\vspace{-2pt}
\label{capture-hpos}
\end{function}

To illustrate the need for $(\INDENT, t_j)$ versus plain \INDENT, consider the following Java method fragment where the first statement is not indented from the previous line.

\begin{alltt}\small
public void write(String str)
\vizsp\vizsp\vizsp\vizsp{}throws IOException \{
\vizsp\vizsp\vizsp\vizsp{}\chpos{\mathtt{int}}{\hspace{4mm}} n = 0;
\end{alltt}

\noindent Directive $(indent,\mathtt{public})$ captures the indentation of {\tt int} but plain \INDENT does not. Plain \INDENT would mean indenting 4 spaces from {\tt throws}, the first token on the previous line, incorrectly indenting {\tt int} 8 spaces relative to {\tt public}.

Directive $indent$ is used to approximate nonstandard indentation as in the following fragment.

\begin{alltt}\small
f(100, 
\vizsp\vizsp\vizsp\chpos{\mathtt{0}}{}); 
\end{alltt}

\noindent At the indicated position, the whitespace does not represent alignment or standard 4 space indentation.   As a default for any nonstandard indentation, function {\em capture\_hpos} returns plain \INDENT as an approximation.

When no suitable alignment or indentation token is available, but the current token is aligned with the previous line, training captures the situation with directive $align$:

\begin{alltt}\small
return x +
\vizsp\vizsp\vizsp\vizsp{}y +
\vizsp\vizsp\vizsp\vizsp{}z; // align with first token of previous line
\end{alltt}

\noindent While $(align,\mathtt{y})$ is valid, that directive is not available because of limitations in how \HPOS directives identify previous tokens, as discussed next.

\subsection{How Directives Refer to Earlier Tokens}

The manner in which training identifies previous tokens for \HPOS directives is critical to successfully formatting documents and is one of the key contributions of this paper.  The goal is to define a ``token locator'' that is as general as possible but that uses the least specific information. The more general the locator, the more previous tokens directives can identify. But, the more specific the locator, the less applicable it is in other contexts.  Consider the indicated positions within the following Java fragments where $align$ directives  must identify the first token of previous function arguments.

\vspace{2mm}

\begin{tabular}{ c c c }
\begin{minipage}[t]{.25\linewidth}
\begin{alltt}\small
f(x, 
\vizsp\vizsp\chpos{\mathtt{y}}{a})
\end{alltt}
\end{minipage} & 
\begin{minipage}[t]{.3\linewidth}
\begin{alltt}\small
f(x+1, 
\vizsp\vizsp\chpos{\mathtt{y}}{b})
\end{alltt}
\end{minipage} & 
\begin{minipage}[t]{.25\linewidth}
\begin{alltt}\small
f(x+1, 
\vizsp\vizsp{}y,
\vizsp\vizsp\chpos{\mathtt{-}}{c}z)
\end{alltt}
\end{minipage}\\
\end{tabular}

\vspace{2mm}

The absolute token index within a document is a completely general locator but is so specific as to be inapplicable to other documents or even other positions within the same document.  For example, all positions \pos{a}, \pos{b}, and \pos{c} could use a single formatting directive, $(align, i)$, but {\tt x}'s absolute index, $i$, is valid only for a function call at that specific location.

The model also cannot use a relative token index referring backwards. While still fully general, such a locator is still too specific to a particular phrase. At position \pos{a}, token {\tt x} is at delta 2, but at position \pos{b}, {\tt x} is at delta 4. Given argument expressions of arbitrary size, no single token index delta is possible and so such deltas would not be widely applicable. Because the delta values are different, the model could not use a single formatting directive to mean ``align with previous argument.'' The more specific the token locator, the more specific the context information in the feature vectors needs to be, which in turn, requires larger corpora (see Section \ref{features}).

We have designed a token locator mechanism that strikes a balance between generality and applicability.  Not every previous token is reachable but the mechanism yields a single locator for {\tt x} from all three positions above and has proven widely applicable in our experiments.  The idea is to pop up into the parse tree and then back down to the token of interest, {\tt x}, yielding a locator with two components: A path length to an ancestor node and a child index whose subtree's leftmost leaf is the target token. This alters formatting directives relative to previous tokens to be: $(\_,ancestor \Delta, child)$.

\begin{figure}
\noindent \scalebox{.67}{\includegraphics{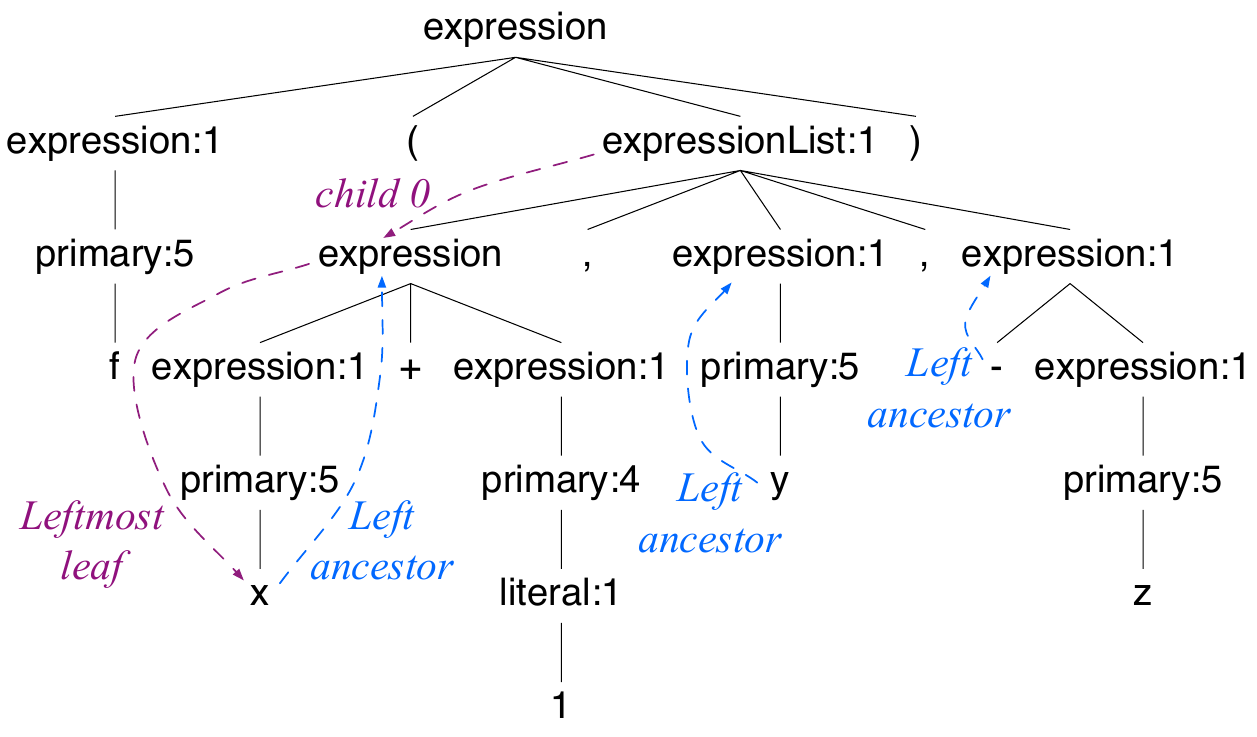}}
\vspace{-8mm}
\caption{\small Parse tree for {\tt f(x+1,y,-z)}. Node {\em rule:n} in the tree indicates the grammar rule and alternative production number used to match the subtree phrase.}
\label{args}
\end{figure}

Unfortunately, training can make no assumptions about the structure of the provided grammar and, thus, parse-tree structure. So, training at $t_i$ involves climbing upwards in the tree looking for a suitable ancestor. To avoid the same issues with overly-specific elements that token indexes have, the path length is relative to what we call the {\em earliest left ancestor} as shown in the parse tree in Figure \ref{args} for {\tt f(x+1,y,-z)}.

The {\em earliest left ancestor} (or just {\em left ancestor}) is the oldest ancestor of $t$ whose leftmost leaf is $t$, and identifies the largest phrase that starts with $t$. (For the special case where $t$ has no such ancestor, we define left ancestor to be $t$'s parent.) It attempts to answer ``what kind of thing  we are looking at.'' For example, the left ancestor computed from the left edge of an arbitrarily-complex expression always refers to the root of the entire expression. In this case, the left ancestors of {\tt x}, {\tt y}, and {\tt z} are siblings, thus, normalizing leaves at three different depths to a common level. The token locator in a directive for {\tt x} in {\tt f(x+1,y,-z)} from both {\tt y} and {\tt z} is $(\_,ancestor \Delta, child)$ = $(\_,1,0)$, meaning jump up 1 level from the left ancestor and down to the leftmost leaf of the ancestor's child 0.  

\begin{figure}
\scalebox{.75}{\includegraphics{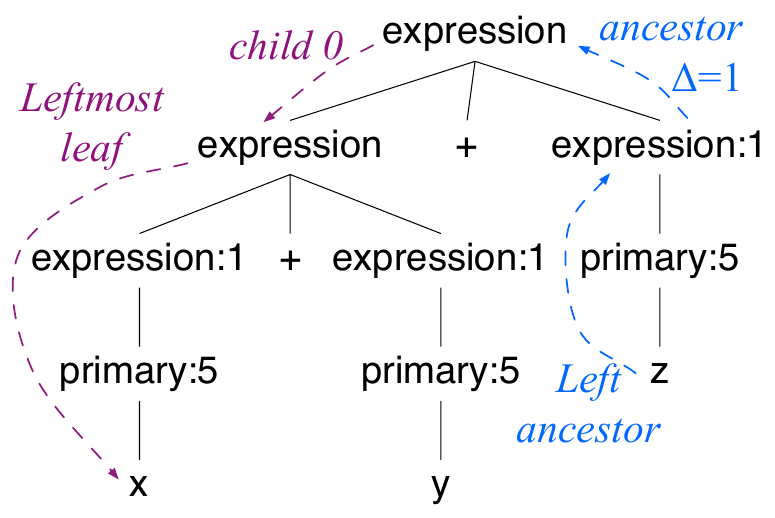}}
\vspace{-5mm}
\caption{\small Parse tree for {\tt x+y+z;}.}
\label{ret-expr}
\end{figure}

The use of the left ancestor and the ancestor's leftmost leaf is critical because it provides a normalization factor among dissimilar parse trees about which training has no inherent structural information.   Unfortunately, some tokens are unreachable using purely leftmost leaves.  Consider the {\tt return x+y+z;} example from the previous  section and one possible parse tree for it in Figure \ref{ret-expr}. Leaf {\tt y} is unreachable as part of formatting directives for {\tt z} because {\tt y} is not a leftmost leaf of an ancestor of {\tt z}. Function {\em capture\_hpos} must either align or indent relative to {\tt x} or fall back on the plain \ALIGN and \INDENT.

The opposite situation can also occur, where a given token is unintentionally aligned with or indented from multiple tokens.  In this case, training chooses the directive with the smallest $ancestor\Delta$, with ties going to indentation. 

And, finally, there could be multiple suitable tokens that share a common ancestor but with different child indexes. For example, if all arguments of {\tt f(x+1,y,-z)} are aligned, the parse tree in Figure \ref{args} shows that $(align,1,0)$ is suitable to align {\tt y} and both $(align,1,0)$ and $(align,1,2)$ could align argument {\tt -z}. Ideally, the formatter would align all function arguments with the same directive to reduce uncertainty in the classifier function (Section \ref{sec:prediction}) so training chooses $(align,1,0)$ for both function arguments.

The formatting directives capture whitespace in between tokens but training must also record the context in which those directives are valid, as we discuss next.

\subsection{Token Context---Feature Vectors}
\label{features}

For each token present in the corpus, training computes an exemplar that associates a context with a \WS and \HPOS formatting-directive: $(X,w,h)$.  Each context has several features combined into a $m$-dimensional feature vector, $X$.  The context information captured by the features must be specific enough to distinguish between language phrases requiring different formatting but not so specific that classifier functions cannot recognize any contexts during formatting.  The shorter the feature vector, the more situations in which each exemplar applies.  Adding more features also has the potential to confuse the classifier.

Through a combination of intuition and exhaustive experimentation, we have arrived at a small set of features that perform well.  There are 22 context features computed during training for each token, but \WS prediction uses only $11$ of them and \HPOS uses $17$.  (The classifier function knows which subset to use.) The feature set likely characterises the context needs of the languages we tested during development to some degree, but the features appear to generalize well (Section \ref{generalize}).

Before diving into the feature details, it is worth describing how we arrived at these 21 features and how they affect formatter precision and generality. We initially thought that a sliding window of, say, four tokens would be sufficient context to make the majority of formatting decisions.  For example, the context for $\cdots${\tt x=\chpos{\tt 1}{}*}$\cdots$ would simply be the token types of the surrounding tokens: $X$=[{\em id},{\tt =},{\em int\_literal},{\tt *}].  The surrounding tokens provide useful but highly-specific information that does not generalize well.  Upon seeing this exact sequence during formatting, the classifier function would find an exact match for $X$ in the model and predict the associated formatting directive.   But, the classifier would not match context $\cdots${\tt x=\chpos{\tt y}{}+}$\cdots$ to the same $X$, despite having the same formatting needs.  

The more unique the context, the more specific the formatter can be.  Imagine a context for token $t_i$ defined as the 20-token window surrounding each $t_i$. Each context derived from the corpus would likely be unique and the model would hold a formatting directive specific to each token position of every file. A formatter working from this model could reproduce with high precision a very similar unknown file.  The trade-off to such precision is poor generality because the model has ``overfit'' the training data.  The classifier would likely find no exact matches for many contexts, forcing it to predict directives from poorly-matched exemplars.

\begin{figure}
\begin{small}
\begin{tabular}[htp]{c|r|c|c}
Corpus & $N$ tokens & Unique \WS & Unique \HPOS \\
\hline
antlr & 19,692 & 3.0\% & 4.7\%\\
java & 42,032 & 3.9\% & 17.4\%\\
java8 & 42,032 & 3.4\% & 7.5\%\\
java\_guava & 499,029 & 0.8\% & 8.1\%\\
sqlite & 14,758 & 8.4\% & 30.8\%\\
tsql & 14,782 & 7.5\% & 17.9\%\\
\end{tabular}
\end{small}
\vspace{-4mm}
\caption{\small Percentage of unique context vectors in corpora.}
\label{unique-X}
\end{figure}

To get a more general model, context vectors use at most two exact token type but lots of context information from the parse tree (details below).  The parse tree provides information about the {\em kind} of phrase surrounding a token position rather than the specific tokens, which is exactly what is needed to achieve good generality. For example, rather than relying solely on the exact tokens after a {\tt =} token, it is more general to capture the fact that those tokens begin an expression.  A useful metric is the percentage of unique context vectors, which we counted for several corpora and show in Figure \ref{unique-X}. Given the features described below, there are very few unique context for \WS decisions (a few \%). The contexts for \HPOS decisions, however, often have many more unique contexts because \WS uses 11-vectors and \HPOS uses 17-vectors. E.g., our reasonably clean SQL corpus has 31\% and 18\% unique \HPOS vectors when trained using SQLite and TSQL grammars, respectively.

For generality, the fewer unique contexts the better, as long as the formatter performs well. At the extreme, a model with just one $X$ context would perform very poorly because all exemplars would be of the form $(X,\_,\_)$. The formatting directive appearing most often in the corpus would be the sole directive returned by the classifier function for any $X$. The optimal model would have the fewest unique contexts but all exemplars with the same context having identical formatting directives.  For our corpora, we found that a majority of unique contexts for \WS and almost all unique contexts for \HPOS predict a single formatting directive, as shown in Figure \ref{ambig-directives}. For example, 57.1\% of the unique {\tt antlr} corpus contexts are associated with just one \WS directive and 95.7\% of the unique contexts predict one \HPOS directive. The higher the ambiguity associated with a single context vector, the higher the uncertainty when predicting formatting decisions during formatting. 

The {\tt guava} corpus stands out as having very few unique contexts for \WS and among the fewest for \HPOS.   This gives a hint that the corpus might be much larger than necessary because the other Java corpora are much smaller and yield good formatting results. Figure \ref{corpus-size} shows the effect of corpus size on classifier error rates. The error rate flattens out after training on about 10 to 15 corpus files.

In short, few unique contexts gives an indication of the potential for generality and few ambiguous decisions gives an indication of the model's potential for accuracy. These numbers do not tell the entire story because some contexts are used more frequently than others and those might all predict single directives. Further, a context associated with multiple directives could be 99\% one specific directive. 

\begin{figure} 
\renewcommand\theadgape{\Gape[4pt]}
\renewcommand\cellgape{\Gape[4pt]}
\begin{small}
\begin{tabular}[htp]{c|r|c}
Corpus & \thead{\small{}Ambiguous\\ \small{}\WS directives} & \thead{\small{}Ambiguous\\ \small{}\HPOS directives}\\
\hline
antlr & 42.9\% & 4.3\% \\
java & 29.8\% & 1.7\%\\
java8 & 31.6\% & 3.2\%\\
java\_guava & 23.5\% & 2.8\%\\
\hline
sqlite\_noisy & 43.5\% & 5.0\%\\
sqlite & 24.8\% & 5.5\%\\
\hline
tsql\_noisy & 40.7\% & 6.3\%\\
tsql & 29.3\% & 6.2\%\\
\end{tabular}
\end{small}
\vspace{-5mm}
\caption{\small Percentage of unique context vectors in corpora associated with $>1$ formatting directive.}
\label{ambig-directives}
\end{figure}

With this perspective in mind, we turn to the details of the individual features.  The \WS and \HPOS decisions use a different subset of features but we present all features computed during training together, broken into three logical subsets.

\subsubsection{Token type and matching token features}
\label{sec:matching-tokens}

At token index $i$ within each document, context feature-vector $X_i$ contains the following features related to previous tokens in the same document.

\begin{enumerate}
\item $t_{i-1}$, token type of previous token
\item $t_i$, token type of current token
\item Is $t_{i-1}$ the first token on a line?
\item Is paired token for $t_i$ the {\em first} on a line?
\item Is paired token for $t_i$ the {\em last} on a line?%
\setcounter{featurecounter}{\theenumi}
\end{enumerate}

\noindent Feature \#3 allows the model to distinguish between the following two different ANTLR grammar rule styles at \pos{a}, when $t_i$={\tt DIGIT}, using two different contexts.

\vspace{2mm}

\noindent \begin{tabular}{ l  l  }
\begin{minipage}[t]{.5\linewidth}
\begin{alltt}\small
DECIMAL\vizsp:\vizsp{}\!\!\chpos{\tt DI}{a}GIT+\vizsp\chpos{\tt \!\!\!;}{b}
\,
\,
\end{alltt}
\end{minipage} &
\begin{minipage}[t]{.5\linewidth}
\begin{alltt}\small
DECIMAL
\vizsp\vizsp\vizsp\vizsp:\vizsp\vizsp\vizsp\!\!\chpos{\tt DI}{a}GIT+
\vizsp\vizsp\vizsp\vizsp\chpos{\tt \!\!\!;}{b}
\end{alltt}
\end{minipage}\\ 
\end{tabular}

\vspace{2mm}

\noindent Exemplars for the two cases are:

\vspace{2mm}
($X$=[{\tt :}, {\em RULEREF}, false, \ldots], $w$=$(sp,1)$, ~$h$=\NONE)\\
\indent ($X'$=[{\tt :}, {\em RULEREF}, true, \ldots], ~$w'$=$(sp,3)$, $h'$=\NONE)
\vspace{2mm}

\noindent where {\em RULEREF} is $type({\tt DIGIT})$, the token type of rule reference {\tt DIGIT} from the ANTLR meta-grammar. Without feature \#3, there would be a single context associated with two different formatting directives.

Features \#4 and \#5 yield different contexts for common situations related to paired symbols, such as {\tt \{} and {\tt \}}, that require different formatting. For example, at position \pos{b}, the model knows that {\tt :} is the paired previous symbol for {\tt ;} (details below) and distinguishes between the styles.  On the left, {\tt :} is not the first token on a line whereas {\tt :} does start the line for the case on the right, giving two different exemplars:

\vspace{2mm}

\cut{ 
($X$=[{\tt +}, {\tt ;}, false, false, true], $w$=$(sp,1)$, ~$h$=\NONE)\\
\indent ($X'$=[{\tt +}, {\tt ;}, false, true, false], $w'$=$(nl,1)$, $h'$=(\ALIGN,{\tt :}))
}
($X$=[\ldots, false, false], $w$=$(sp,1)$, ~$h$=\NONE)\\
\indent ($X'$=[\ldots, true, false], $w'$=$(nl,1)$, $h'$=(\ALIGN,{\tt :}))

\vspace{2mm}

\noindent  Those features also allow the model to distinguish between the first two following Java cases where the paired symbol for {\tt \}} is sometimes not at the end of the line in short methods.

\vspace{2mm}

\hspace{-7mm} \begin{tabular}{ l | l | l}
\begin{minipage}[t]{.36\linewidth}
\begin{alltt}\small
void reset() \{x=0;\}
\,
\,
\end{alltt}
\end{minipage} &
\begin{minipage}[t]{.26\linewidth}
\begin{alltt}\small
void reset() \{
    x=0;
\}
\end{alltt}
\end{minipage} &
\begin{minipage}[t]{.05\linewidth}
\begin{alltt}\small
void reset() \{
    x=0;\}
\end{alltt}
\end{minipage}\\
\end{tabular}

\vspace{2mm}

\noindent Without features \#4-\#5, the formatter would yield the third.

Determining the set of paired symbols is nontrivial, given that the training can make no assumptions about the language it is formatting.  We designed an algorithm, {\em pairs} in Function \ref{pairs}, that analyzes the parse trees for all documents in the corpus and computes plausible token pairs for every non-leaf node (grammar rule production) encountered. The algorithm relies on the idea that paired tokens are token literals, occur as siblings, and are not repeated siblings. Grammar authors also do not split paired token references across productions. Instead, authors write productions such as these ANTLR rules for Java:

\begin{alltt}\small
expr : ID '[' expr ']' | ... ;
type : ID '<' ID (',' ID)* '>' | ... ;
\end{alltt}

\noindent that yield subtrees with the square and angle brackets as direct children of the relevant production. Repeated tokens are not plausible pair elements so the commas in a generic Java type list, as in {\tt T<A,B,C>}, would not appear in pairs associated with rule {\tt type}.  A single subtree in the corpus with repeated commas as children of a {\tt type} node would remove comma from all pairs associated with rule {\tt type}. Further details are available in Function \ref{pairs} (source {\small CollectTokenPairs.java}). The algorithm neatly identifies pairs such as ({\tt ?}, {\tt :}) and ({\tt [}, {\tt ]}), and ({\tt (}, {\tt )}) for Java expressions and ({\tt enum}, {\tt \}}), ({\tt enum}, {\tt \{}), and ({\tt \{}, {\tt \}}) for enumerated type declarations. During formatting, {\em paired} (Function \ref{paired}) returns the paired symbols for $t_i$.

\setlength{\algomargin}{3pt}
\SetAlCapSkip{-10pt}
\begin{function}[]
\SetAlgorithmName{Alg.}{List of Algorithms}
\SetAlgoSkip{}
\SetInd{.5em}{.5em}
\TitleOfAlgo{$pairs$(Corpus $D$) $\rightarrow$ map node $\mapsto$ set{\tt <}($s,t$){\tt >}}
{\em pairs} := map of node $\mapsto$ set\verb|<|tuples\verb|>|\;
{\em repeats} := map of node $\mapsto$ set\verb|<|token types\verb|>|\;
\ForEach{$d \in D$}{
  \ForEach{non-leaf node $r$ in parse($d$)}{
    {\em literals} := $\{ t \mid parent(t)=r, t \text{ is literal token}\}$\;
    add $\{(t_i,t_j) \mid i<j \,\forall\, t_i,t_j \in literals\}$ to {\em pairs}[$r$]\;
    add $\{ t_i \mid \exists\, t_i=t_j, i\neq{}j\}$ to  {\em repeats}[$r$]\;
  }
}
delete pair $(t_i,t_j) \in pairs[r]$ if $t_i \text{ or } t_j \in repeats[r] \,\forall\, r$\;
\Return{pairs}\;
\vspace{-2pt}
\label{pairs}
\end{function}

\setlength{\algomargin}{3pt}
\SetAlCapSkip{-10pt}
\begin{function}[]
\SetAlgorithmName{Alg.}{List of Algorithms}
\SetAlgoSkip{}
\SetInd{.5em}{.5em}
\TitleOfAlgo{$paired$({\em pairs}, token $t_i$) $\rightarrow$ $t'$}
{\em mypairs} := {\em pairs}[$parent(t)$]\;
{\em viable} := $\{s \mid (s,t) \in {\textit mypairs}, s \in {\textit siblings}(t) \}$\;
\lIf{$|$viable$|=1$}{{\em ttype} := {\em viable}[0]\;}
\lElseIf{$\exists (s,t) |~ s,t$ are common pairs}{{\em ttype} := $s$\;}
\lElseIf{$\exists (s,t) |~ s,t$ are single-char literals}{{\em ttype} := $s$\;}
\lElse{{\em ttype} := {\em viable}[0]; // choose first if still ambiguous\\}
{\em matching} := $[ t_j \mid t_j=ttype, j<i, \,\forall\, t_j \in {\textit siblings}(t_i)]$\;
\Return{last(matching)}\;
\vspace{-2pt}
\label{paired}
\end{function}

\subsubsection{List membership features}
\label{sec:lists}

Most computer languages have lists of repeated elements separated or terminated by a token literal, such as statement lists, formal parameter lists, and table column lists. The next group of features indicates whether $t_i$ is a component of a list construct and whether or not that list is split across multiple lines (``oversize'').

\begin{enumerate}
\setcounter{enumi}{\value{featurecounter}}
\item Is $\mathit{leftancestor}(t_i)$ a component of an oversize list?
\item $\mathit{leftancestor}(t_i)$ component type within list from\\
\{{\em prefix token, first member, first separator, member,\\
separator, suffix token}\}
\setcounter{featurecounter}{\theenumi}
\end{enumerate}

\noindent With these two features, context vectors capture not only two different overall styles for short and oversize lists but how the various elements are formatted within those two kinds of lists. Here is a sample oversize Java formal parameter list annotated with list component types:
\begin{center}
\scalebox{.52}{\includegraphics{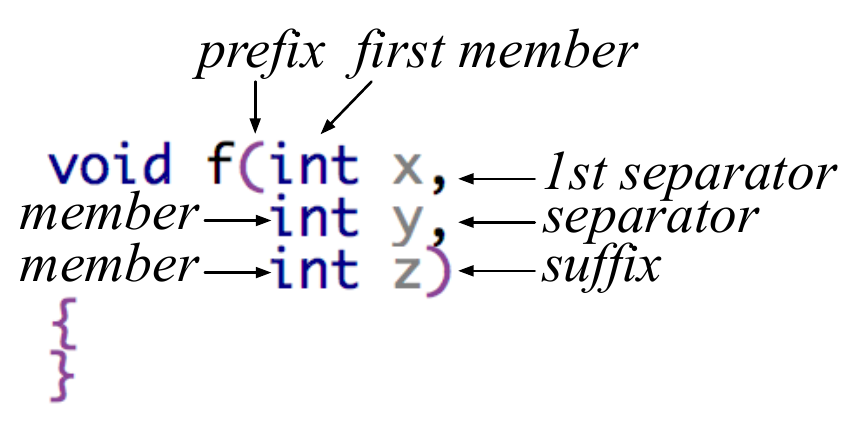}}
\end{center}
\noindent Only the first member of a list is differentiated; all other members are labeled as just plain members because their formatting is typically the same.  The exemplars would be:

\vspace{2mm}

($X$=[\ldots, true, {\em prefix}], $w$=\NONE, $h$=\NONE)\\
\indent ($X$=[\ldots, true, {\em first member}], $w$=\NONE, $h$=\NONE)\\
\indent ($X$=[\ldots, true, {\em first separator}], $w$=\NONE, $h$=\NONE)\\
\indent ($X$=[\ldots, true, {\em member}], $w$=$(\NL,1)$, $h$=$(\ALIGN, \text{first arg})$)\\
\indent ($X$=[\ldots, true, {\em separator}], $w$=\NONE, $h$=\NONE)\\
\indent ($X$=[\ldots, true, {\em member}], $w$=$(\NL,1)$, $h$=$(\ALIGN, \text{first arg})$)\\
\indent ($X$=[\ldots, true, {\em suffix}], $w$=\NONE, $h$=\NONE)

\vspace{2mm}

Even for short lists on one line, being able to differentiate between list components lets training capture different but equally valid styles. For example, some ANTLR grammar authors write short  parenthesized subrules like {\tt (ID|INT|FLOAT)} but some write {\tt (ID | INT | FLOAT)}.

\cut{
\vspace{2mm}

\noindent \begin{tabular}{ l  l  l }
\begin{minipage}[t]{.3\linewidth}
\begin{alltt}\small
(   ID
|   INT
|   FLOAT
)
\end{alltt}
\end{minipage} &
\begin{minipage}[t]{.25\linewidth}
\begin{alltt}\small
void f(
    int x,
    int y
)
\end{alltt}
\end{minipage} &
\begin{minipage}[t]{.3\linewidth}
\begin{alltt}\small
SELECT
    Client\_ID
    , SourceDB
    , Beta \ldots
\end{alltt}
\end{minipage}\\ 
\end{tabular}

\vspace{2mm}
}

\begin{figure}
\scalebox{.6}{\includegraphics{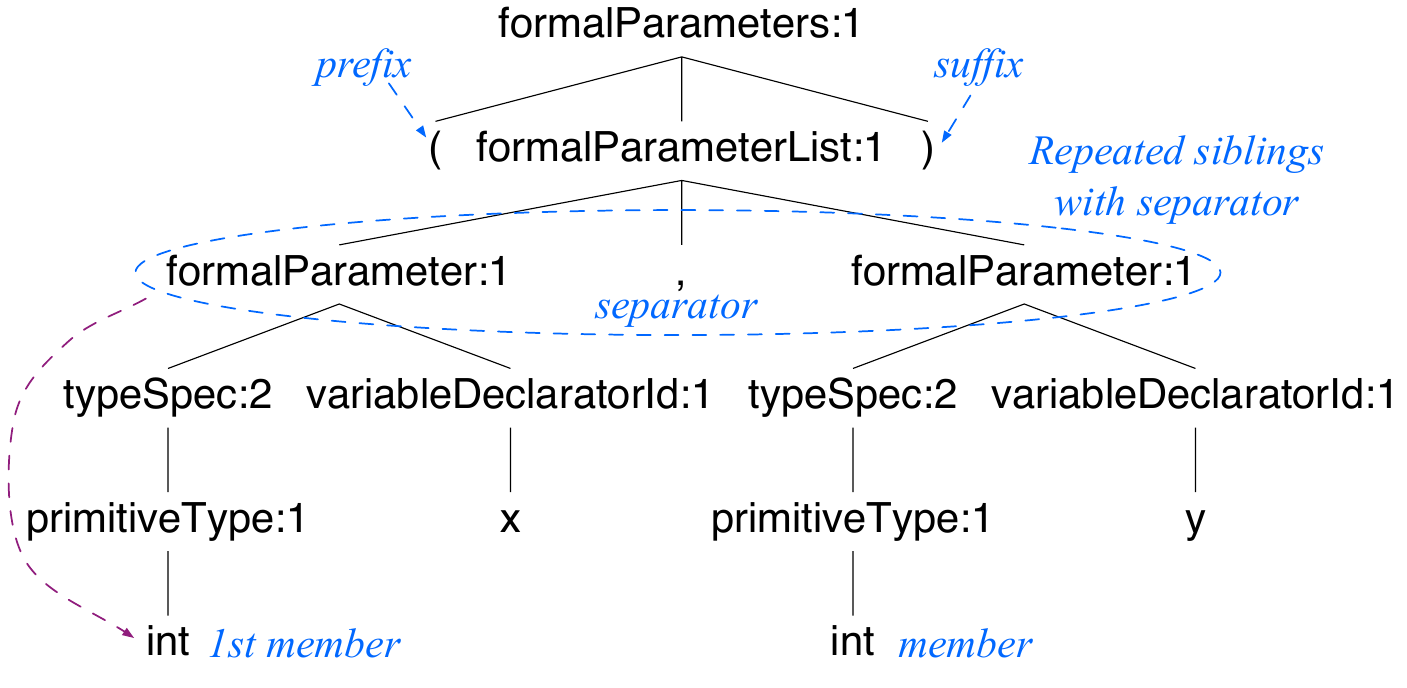}}
\vspace{-8mm}
\caption{\small Formal Args Parse Tree {\tt\small void f(int x, int y)}.}
\label{formal-args}
\end{figure}

As with identifying token pairs, \codebuff must identify the constituent components of lists without making assumptions about grammars that hinder generalization.  The intuition is that  lists are repeated sibling subtrees with a single token literal between the 1st and 2nd repeated sibling, as shown in Figure \ref{formal-args}. Repeated subtrees without separators are not considered lists.  Training performs a preprocessing pass over the parse tree for each document, tagging the tokens identified as list components with values for features \#6- \#7. Tokens starting list members are identified as the leftmost leaves of repeated siblings ({\small\tt formalParameter} in Figure \ref{formal-args}). Prefix and suffix components are the tokens immediately to the left and right of list members but only if they share a common parent.

The training preprocessing pass also collects statistics about the distribution of list text lengths (without whitespace) of regular and oversize lists. Regular and oversize list lengths are tracked per $(r,c,sep)$ combination for rule subtree root type $r$, child node type $c$, and separator token type $sep$; e.g., $(r,c,sep)$=$(${\tt\small formalParameterList}, {\tt\small formalParameter},`{\tt ,}'$)$ in Figure \ref{formal-args}. The separator is part of the tuple so that expressions can distinguish between different operators such as {\tt =} and {\tt *}. Children of binary and ternary operator subtrees satisfy the conditions for being a list, with the operator as  separator token(s).  For each $(r,c,sep)$ combination, training tracks the number of those lists and the median list length, $(r,c,sep) \mapsto (n, median)$.  

\subsubsection{Identifying oversize lists during formatting}

As with training, the formatter performs a preprocessing pass to identify the tokens of list phrases. 
Whereas training identifies oversize lists simply as those split across lines, formatting sees documents with all whitespace squeezed out.  For each $(r,c,sep)$ encountered during the preprocessing pass, the formatter consults a mini-classifier to predict whether that list is oversize or not based upon the list string length, $ll$. The mini-classifier compares the mean-squared-distance of $ll$ to the median for regular lists and the median for oversize (big) lists and then adjusts those distances according to the likelihood of regular vs oversize lists.  The {\em a priori} likelihood that a list is regular is $p(reg)=n_{reg}/(n_{reg}+n_{big})$, giving an adjusted distance to the regular type list as: $dist_{reg} = (ll - median_{reg})^2 * (1 - p(reg))$. The distance for oversize lists is analogous. 

When a list length is somewhere between the two medians, the relative likelihoods of occurrence shift the balance. When there are roughly equal numbers of regular and oversize lists, the likelihood term effectively drops out, giving just mean-squared-distance as the mini-classifier criterion. At the extreme, when all $(r,c,sep)$ lists are big, $p(big)=1$, forcing $dist_{big}$ to 0 and, thus, always predicting oversize.   

When a single token $t_i$ is a member of multiple lists, training and formatting associate $t_i$  with the longest list subphrase because that yields the best formatting, as evaluated manually across the corpora.  For example, the expressions within a Java function call argument list are often themselves lists. In {\tt f($e_1$,...,a+b)}, token {\tt a} is both a sibling of {\tt f}'s argument list but also the first sibling of expression {\tt a+b}, which is also a list. Training and formatting  identify {\tt a} as being part of the larger argument list rather than the smaller {\tt a+b}.  This choice ensures that oversize lists are properly split.  Consider the opposite choice where {\tt a} is associated with list {\tt a+b}. In an oversize argument list, the formatter would not inject a newline before {\tt a}, yielding poor results:

\begin{alltt}\small
f(\(e\sb{1}\),
  ..., a+b)
\end{alltt}

\noindent  Because list membership identification occurs in a top-down parse-tree pass, associating tokens with the largest construct is a matter of latching the first list association discovered.

\subsubsection{Parse-tree context features}

The final features provide parse-tree context information:

\begin{enumerate}\small
\setcounter{enumi}{\value{featurecounter}}
\item $\mathit{childindex}(t_i)$ 
\item $\mathit{rightancestor}(t_{i-1})$
\item $\mathit{leftancestor}(t_i)$
\item $\mathit{childindex}(\mathit{leftancestor}(t_i))$
\item $\mathit{parent}_1(\mathit{leftancestor}(t_i))$
\item $\mathit{childindex}(parent_1(\mathit{leftancestor}(t_i)))$
\item $\mathit{parent}_2(\mathit{leftancestor}(t_i))$
\item $\mathit{childindex}(\mathit{parent}_2(\mathit{leftancestor}(t_i)))$
\item $\mathit{parent}_3(\mathit{leftancestor}(t_i))$
\item $\mathit{childindex}(\mathit{parent}_3(\mathit{leftancestor}(t_i)))$
\item $\mathit{parent}_4(\mathit{leftancestor}(t_i))$
\item $\mathit{childindex}(\mathit{parent}_4(\mathit{leftancestor}(t_i)))$
\item $\mathit{parent}_5(\mathit{leftancestor}(t_i))$
\item $\mathit{childindex}(\mathit{parent}_5(\mathit{leftancestor}(t_i)))$
\end{enumerate}

\noindent Here the $\mathit{childindex}(p)$ is the 0-based index of node $p$ among children of $parent(p)$, $\mathit{childindex}(t_i)$ is shorthand for $\mathit{childindex}(\mathit{leaf}(t_i))$, and $\mathit{leaf}(t_i)$ is the leaf node associated with $t_i$. Function $\mathit{childindex}(p)$ has a special case when $p$ is a repeated sibling. If $p$ is the first element, $\mathit{childindex}(p)$ is the actual child index of $p$ within the children of $\mathit{parent}(p)$ but is special marker * for all other repeated siblings. The purpose is to avoided over-specializing the context vectors to improve generality.  These features also use function $\mathit{parent}_i(p)$, which is the \ith\ parent of $p$; $\mathit{parent}_1(p)$ is synonymous with the direct parent $parent(p)$.

The child index of $t_i$, feature \#8, gives the necessary context information to distinguish the alignment token between the following two ANTLR lexical rules at the semicolon.

\vspace{2mm}

\noindent \begin{tabular}{ c c }
\begin{minipage}[t]{.45\linewidth}
\begin{alltt}\small
BooleanLiteral
    :   'true'  
    |   'false' 
    ;   
\end{alltt}
\end{minipage} &
\begin{minipage}[t]{.55\linewidth}
\begin{alltt}\small
fragment
DIGIT
    :   [0-9]
    ;
\end{alltt}
\end{minipage}\\ 
\end{tabular}

\vspace{2mm}

\noindent On the left, the {\tt ;} token is child index 3 but 4 on the right, yielding different contexts, $X$ and $X'$, to support different alignment directives for the two cases. Training collects exemplars $(X, (\ALIGN,0,1))$ and $(X', (\ALIGN,0,2))$, which aligns {\tt ;} with the colon in both cases.

Next, features $\mathit{rightancestor}(t_{i-1})$ and $\mathit{leftancestor}(t_i)$ describe what phrase precedes $t_i$ and what phrase $t_i$ starts. The $\mathit{rightancestor}$ is analogous to $\mathit{leftancestor}$ and is the oldest ancestor of $t_i$ whose rightmost leaf is $t_i$ (or $\mathit{parent}(t_i)$ if there is no such ancestor).  For example, at $t_i$={\tt y} in {\tt x=1; y=2;} the right ancestor of $t_{i-1}$ and the left ancestor of $t_i$ are both ``statement'' subtree roots.

\begin{figure}
\scalebox{.54}{\includegraphics{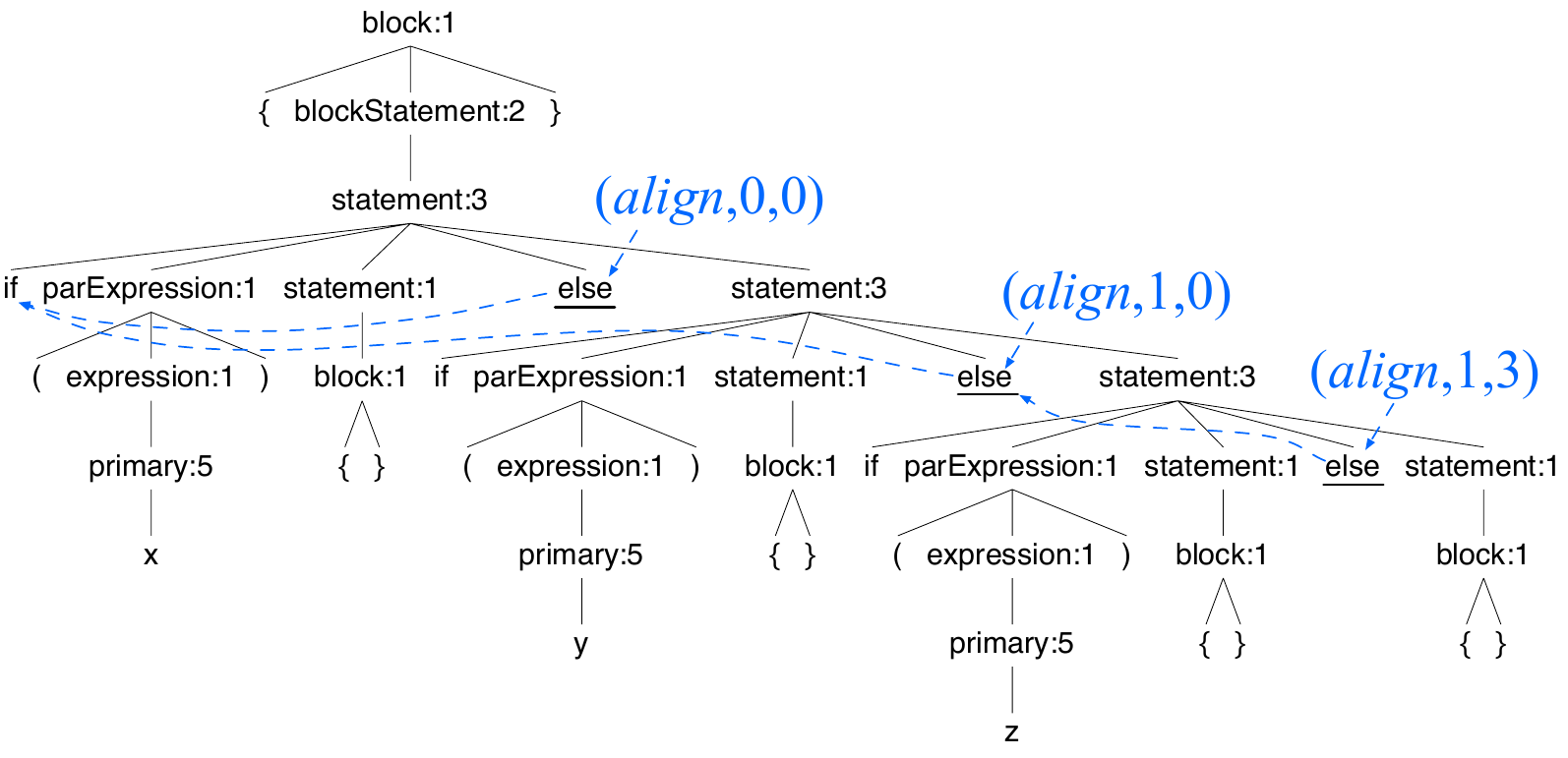}}
\vspace{-5mm}
\caption{\small Alignment directives for nested {\tt if}-{\tt else} statements.}
\label{elses}
\end{figure}

Finally, the parent and child index features capture context information about highly nested constructs, such as:

\begin{alltt}\small
if ( x ) \{ \}
else if ( y ) \{ \}
else if ( z ) \{ \}
else \{ \}
\end{alltt}

\noindent Each {\tt else} token requires a different formatting directive for alignment, as shown in Figure \ref{elses}; e.g., $(\ALIGN,1,3)$ means ``jump up 1 level from $\mathit{leftancestor}(t_i)$ and align with leftmost leaf of child 3 (token {\tt else}).'' To distinguish the cases, the context vectors must be different. Therefore, training collects these partial vectors with features \#10-15:

\vspace{2mm}

$X$=[\ldots, {\em stat}, 0, {\em blockStat}, *, {\tt block}, 0, \ldots]\\
\indent $X$=[\ldots, {\em stat}, *, {\em stat}, 0, {\em blockStat}, *, \ldots]\\
\indent $X$=[\ldots, {\em stat}, *, {\em stat}, *, {\em stat}, 0, \ldots]

\vspace{2mm}

\noindent where {\em stat} abbreviates {\small\tt statement:3} and {\em blockStat} abbreviates {\small\tt blockStatement:2}. All deeper {\tt else} clauses also use directive $(\ALIGN,1,3)$.

\cut{
\todo: model can't distinguish between context of {\tt \}} end of methods:
\begin{alltt}
@Override
T foo() \{
\}
public T bar() \{
\}
\end{alltt}
context predicts align,4,0 and align,2,0 for first and second.
}

Training is complete once the software has computed an exemplar for each token in all corpus files. The formatting model is the collection of those exemplars and an associated classifier that predicts directives given a feature vector. 

\subsection{Predicting Formatting Directives}
\label{sec:prediction}

\codebuff's {\em kNN} classifier uses a fixed $k=11$ (chosen experimentally in Section \ref{empirical}) and an $L_0$ distance function (ratio of number of components that differ to vector length) but with a twist on classic {\em kNN} that accentuates feature vector distances in a nonlinear fashion.  To make predictions, a classic {\em kNN} classifier computes the distance from unknown feature vector $X$ to every $X_j$ vector in the exemplars, $({\bf X}, Y)$, and predicts the category, $y$, occurring most frequently among the $k$ exemplars nearest $X$. 

The classic approach works very well in Euclidean space with quantitative feature vectors but not so well with an $L_0$ distance that measures how similar two code-phrase contexts are. As the $L_0$ distance increases, the similarity of two context vectors drops off dramatically. Changing even one feature, such as earliest left ancestor (kind of phrase), can mean very different contexts.  This quick drop off matters when counting votes within the $k$ nearest $X_j$. At the extreme, there could be one exemplar where $X=X_j$ at distance 0 and 10 exemplars at distance 1.0, the maximum distance. Clearly the one exact match should outweigh 10 that do not match at all, but a classic {\em kNN} uses a simple unweighted count of exemplars per category (10 out of 11 in this case).  Instead of counting the number of exemplars per category, our variation sums $1 - \sqrt[3]{L_0(X,X_j)}$ for each $X_j$ per category. Because distances are in $[0..1]$, the cube root nonlinearly accentuates differences.  Distances of 0 count as weight 1, like the classic {\em kNN}, but distances close to 1.0 count very little towards their associated category.  In practice, we found feature vectors more distant than about 15\% from unknown $X$ to be too dissimilar to count.  Exemplars at distances above this threshold are discarded while collecting the $k$ nearest neighbors.

\cut{
\setlength{\algomargin}{3pt}
\SetAlCapSkip{-10pt}
\begin{function}[]
\SetAlgorithmName{Alg.}{List of Algorithms}
\SetAlgoSkip{}
\SetInd{.5em}{.5em}
\TitleOfAlgo{$predict$($Ctx,f,Y,k$)\vspace{-3pt}}
$N$ = len($Ctx$)\;
$\Delta = []$\;
\For{$i=0,N-1$}{
let $\delta_i = distance(Ctx_i, f)$\;
$\Delta_i = (\delta_i, i)$\;
}
Sort $\Delta$ ascending\;
$votes = \Delta[0:k]$\;
$\sum_i Ctx_i * \delta_i$\;
\Return{$Y_i \leq 1/M$ with most votes}\;
\vspace{-2pt}
\label{parse}
\end{function}
}

The classfier function uses features \#1-\#10, \#12 to make \WS predictions and \#2, \#6-\#21 for \HPOS; \HPOS predictions ignore $X_j$ not associated with tokens starting a line.

\subsection{Formatting a Document}

To format document $d$, the formatter Function \ref{format} first squeezes out all whitespace tokens and line/column information from the tokens of $d$ and then iterates through $d$'s remaining tokens, deciding what whitespace to inject before each token. At each token, the formatter computes a feature vector for that context and asks the model to predict a formatting directive (whereas training examines the whitespace to determine the directive).  The formatter uses the information in the formatting directive to compute the number of newline and space characters to inject.  The formatter treats the directives like bytecode instructions for a simple virtual machine: \{$(\mathit{nl},n)$, $(\mathit{sp},n)$, \NONE, $(\mathit{align,ancestor\Delta,child})$, $(\mathit{indent,ancestor\Delta,child})$, \ALIGN, $\mathit{indent}$\}.

As the formatter emits tokens and injects whitespace, it tracks line and column information so that it can annotate tokens with this information.  Computing features \#3-5 at token $t_i$ relies on line and column information for $t_j$ for some $j<i$.  For example, feature \#3 answers whether $t_{i-1}$ is the first token on the line, which requires line and column information for $t_{i-1}$ and $t_{i-2}$.  Because of this, predicting the whitespace preceding token $t_i$ is a (fast) function of the actions made previously by the formatter. After processing  $t_i$, the file is formatted up to and including $t_i$.
 
Before emitting whitespace in front of token $t_i$, the formatter emits any comments found in the source code. (The ANTLR parser has comments available on a ``hidden channel''.) To get the best output, the formatter needs whitespace in front of comments and this is the one case where the formatter looks at the original file's whitespace. Otherwise, the formatter computes all whitespace generated in between tokens. To ensure single-line comments are followed by a newline, users of \codebuff can specify the token type for single-line comments as a failsafe.

\SetKwFor{ForEach}{foreach}{do}{\vspace{-12pt}}
\setlength{\algomargin}{3pt}
\SetAlCapSkip{-10pt}
\begin{function}[]
\SetAlgorithmName{Alg.}{List of Algorithms}
\SetAlgoSkip{}
\SetInd{.5em}{.5em}
\TitleOfAlgo{{\em format}($F_{D,G}=({\bf X}, W, H, \operatorname{indentSize}), d$)\vspace{-3pt}}
{\em line} := {\em col} := 0\;
$d$ := $d$ with whitespace tokens, line/column info removed\;
\ForEach{$t_i \in d$}{
emit any comments to left of $t_i$\;
$X_i$ := compute context feature vector at $t_i$\;
$ws$ := predict directive using $X_i$ and {\bf X}, $W$\;
{\em newlines} := {\em sp} := 0\;
\lIf{$ws$ = $(nl,n)$}{{\em newlines} := $n$\;} 
\lElseIf{$ws$ = $(sp,n)$}{{\em sp} := $n$\;}
\uIf(// inject newline and align/indent){newlines $>0$}{
  emit $newlines$ `{\tt \textbackslash{}n}' characters\;
  {\em line} += $newlines$; {\em col} := 0\;
  $hpos$ := predict directive using $X_i$ and {\bf X}, $H$\;
  \uIf{$hpos = (\_, ancestor\Delta, child)$}{
    $t_j$ = token relative to $t_i$ at $ancestor\Delta, child$\;
    {\em col} := $t_j$.col\;
    \lIf{$hpos = (indent, \_, \_)$}{{\em col} += $\operatorname{indentSize}$\;}
    emit {\em col} spaces\;
  }
  \Else(// plain align or indent){
    $t_j$ := first token on previous line\;
    {\em col} := $t_j$.col\;
    \lIf{$hpos = indent$}{{\em col} += $indentSize$\;}
    emit {\em col} spaces\;
  }
}
\Else{
  {\em col} += {\em sp}\;
  emit {\em sp} spaces; // inject spaces\\
}
$t_i$.line = {\em line};  ~~~~~~~~// set $t_i$ location\\
$t_i$.col = {\em col}\;
emit $text(t_i)$\;
{\em col} += $len(text(t_i))$
}
\vspace{-2pt}
\label{format}
\end{function}

\section{Empirical results\label{sec:evaluation}\label{sec:experiments}
\label{empirical}}

The primary question when evaluating a code formatter is whether it consistently produces high quality output, and we begin by showing experimentally that \codebuff does so. Next, we investigate the key factors that influence \codebuff's statistical model and, indirectly, formatting quality: the way a grammar describes a language, corpus size/consistency, and parameter $k$ of the {\em kNN} model. We finish with a discussion of \codebuff's complexity and performance.

\subsection{Research Method: Quantifying formatting quality}

We need to accurately quantify code formatter quality without human evaluation. A metric helps to isolate issues with the model (and subsequently improve it) as well as report its efficacy in an objective manner. We propose the following measure. Given corpus $D$ that is perfectly consistently formatted, \codebuff should produce the identity transformation for any document $d \in D$ if trained on a corpus subset $D \setminus \{d\}$. This {\em leave-one-out cross-validation} allows us to use the corpus for both training and for measuring formatter quality. (See Section \ref{sec:validation} for evidence of \codebuff's generality.)  For each document, the distance between original $d$ and formatted $d'$ is an inverse measure of formatting quality.

A naive similarity measure is the edit distance (Levenshtein Distance \cite{Levenshtein}) between $d'$ and $d$, but it is expensive to compute and will over-accentuate minor differences. For example, a single indentation error made by the formatter could mean the entire file is shifted too far to the right, yielding a very high edit distance. A human reviewer would likely consider that a small error, given that the code looks exactly right except for the indentation level. Instead, we quantify the document similarity using the aggregate misclassification rate, in $[0..1]$, for all predictions made while generating $d'$:
\[
error = \frac{n\_ws\_errors+n\_hpos\_errors}{n\_ws\_decisions+n\_hpos\_decisions}
\]
\noindent A misclassification error occurs when the {\em kNN} model predicts a formatting directive for $d'$ at token $t_i$ that differs from the actual formatting found in the original $d$ at $t_i$.   The formatter predicts whitespace for each $t_i$ so $n\_ws\_decisions = |d|=|d'|$, the number of real tokens in $d$. For each $ws=(nl,\_)$ prediction, the formatter predicts $hpos$ so $n\_hpos\_decisions \leq |d|$. An error rate of 0 indicates that $d'$ is identical to $d$ and an error rate of 1 indicates that every prediction made during formatting of $d'$ would yield formatting that differs from that found in $d$.  Formatting directives that differ solely in the number of spaces or in the relative token identifier count as misclassifications; e.g., $(\mathit{sp}, 1) \neq (\mathit{sp}, 2)$ and $(\mathit{align,i,j}) \neq (\mathit{align,i',j'})$. We consider this error rate an acceptable proxy for human opinion, albeit imperfect.

\subsection{Corpora}

We selected three very different languages---ANTLR grammars, Java, and SQL---and used the following corpora (stored in \codebuff's~\cite{codebuff} {\tt corpus} directory).

\begin{itemize}[itemsep=-1mm]
\item {\tt antlr}. A subset of 12 grammars from {\bf ANTLR}'s grammar repository, manually formatted by us.
\item {\tt st}. All 59 {\bf Java} source files for StringTemplate.
\item {\tt guava}. All 511 {\bf Java} source files for Google's Guava.
\item {\tt sql\_noisy}. 36 {\bf SQL} files taken from a {\tt github} repository.\footnote{\tt https://github.com/mmessano/SQL}  The SQL corpus was groomed and truncated so it was acceptable to both SQLite and TSQL grammars.
\item {\tt sql}. The same 36 {\bf SQL} files as formatted using Intellij IDE; some manual formatting interventions were done to fix Intellij formatting errors.
\end{itemize}

\noindent  As part of our examination of grammar invariance (details below), we used two different Java grammars and two different SQL grammars taken from ANTLR's grammar repository:

\begin{itemize}[itemsep=-1mm]
\item {\tt java}. A Java 7 grammar.
\item {\tt java8}. A transcription of the Java 8 language specification into ANTLR format.
\item {\tt sqlite}. A grammar for the SQLite variant of SQL.
\item {\tt tsql}. A grammar for the Transact-SQL variant of SQL.
\end{itemize}

\begin{figure}[htbp]
\begin{center}
\scalebox{.35}{\includegraphics{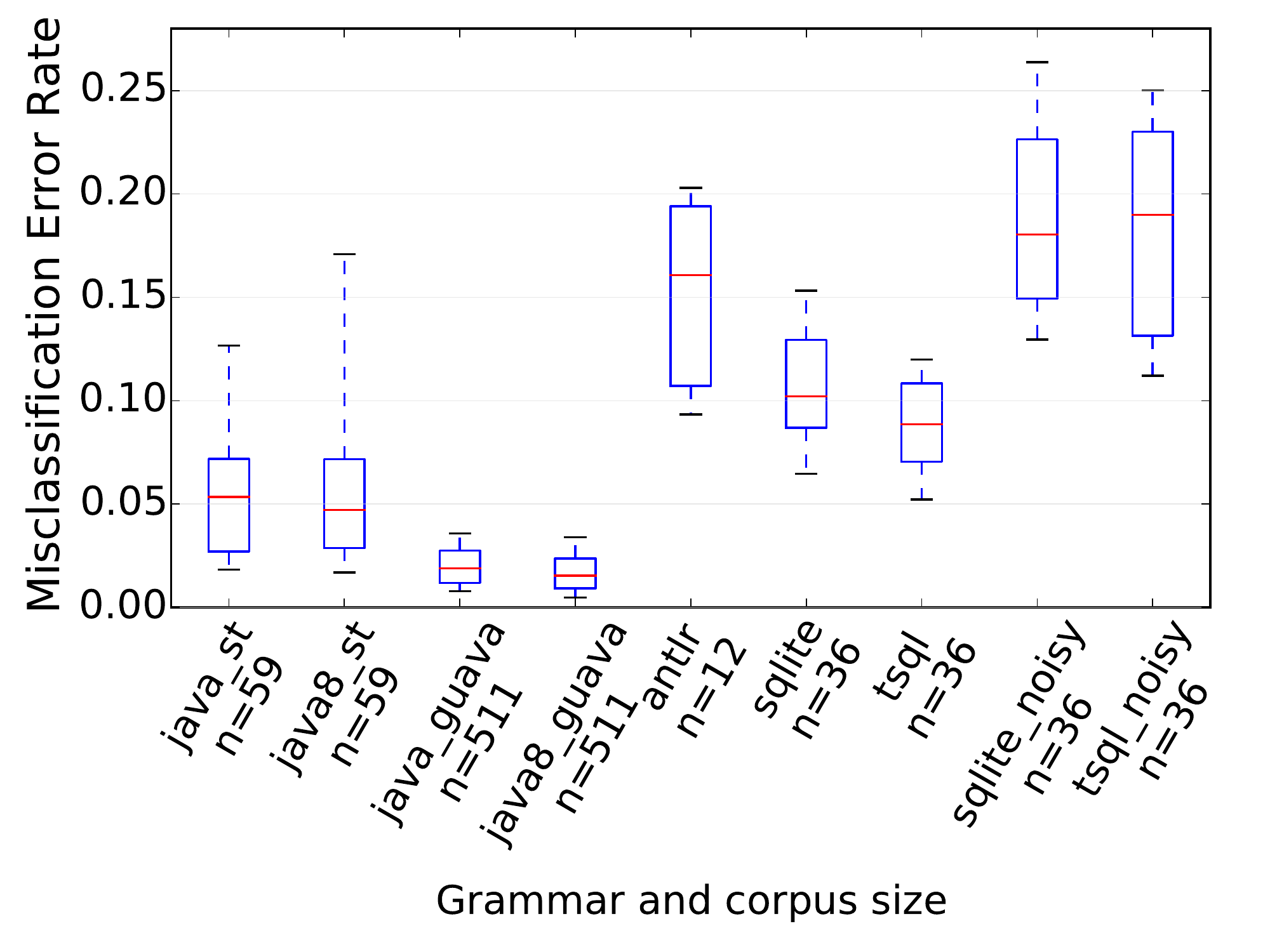}}
\vspace{-7mm}
\caption{\small Standard box-plot of leave-one-out validation error rate between formatted document $d'$ and original $d$.}
\label{leave-one-out}
\end{center}
\end{figure}

\subsection{Formatting quality results}

Our first experiment demonstrates that \codebuff can faithfully reproduce the style found in a consistent corpus. Details to reproduce all results are available in a \texttt{\small README.md}~\cite{codebuff}.  Figure \ref{leave-one-out} shows the formatting error rate, as described above. Lower median error rates correspond with higher-quality formatting, meaning that the formatted files are closer to the original.  Manual inspection of the corpora confirms that consistently-formatted corpora indeed yield better results. For example, median error rates (17\% and 19\%) are higher using the two grammars on the {\tt sql\_noisy} corpus versus the cleaned up {\tt sql} corpus. The {\tt guava} corpus has extremely consistent style because it is enforced programmatically and consequently \codebuff is able to reproduce the style with high accuracy using either Java grammar. The {\tt antlr} corpus results have a high error rate due to some inconsistencies among the grammars but, nonetheless, formatted grammars look good except for a few  overly-long lines.

\subsection{Grammar invariance}

Figure \ref{leave-one-out} also gives a strong hint that \codebuff is {\em grammar invariant}, meaning that training models on a single corpus but with different grammars gives roughly the same formatting results.  For example, the error rates for the {\tt st} corpus trained with {\tt java} and {\tt java8} grammars are roughly the same, indicating that \codebuff's overall error rate (similarity of original/formatted documents) does not change when we swap out the grammar. The same evidence appears for the other corpora and grammars. The overall error rate could hide large variation in the formatting of individual files, however, so we define grammar invariance as a file-by-file comparison of normalized edit distances.

\begin{definition}
Given models $F_{D,G}$ and $F_{D,G'}$ derived from grammars $G$ and $G'$ for a single language, $L(G)=L(G')$, a formatter is {\em grammar invariant} if the following holds for any document $d$: $\mathit{format(F_{D,G},d)} \ominus \mathit{format(F_{D,G'},d) \leq \epsilon}$ for some suitably small {\em normalized edit distance} $\epsilon$.
\label{def:invariance}
\end{definition}

\begin{definition}
Let operator $d_1 \ominus d_2$ be the {\em normalized edit distance} between documents $d_1$ and $d_2$ defined by the Levenshtein Distance \cite{Levenshtein}  divided by $max(len(d_1), len(d_2))$.
\end{definition}

The median edit distances between formatted files (using leave-one-out validation) from 3 corpora provide strong evidence of grammar invariance for Java but less so for SQL:

\begin{itemize}[itemsep=-1mm]
\item 0.001 for {\tt guava} corpus with {\tt java} and {\tt java8} grammars
\item 0.008 for {\tt st} corpus with {\tt java} and {\tt java8} grammars
\item 0.099 for {\tt sql} corpus with {\tt sqlite} and {\tt tsql} grammars
\end{itemize}

\cut{
clean SQLite vs TSQL edit distance info median=0.099117175
Java vs Java8 edit distance info median=0.008094187
Java vs Java8 guava edit distance info median=0.0014512033
}

\noindent The ``average'' difference between {\tt guava} files formatted with different Java grammars is 1 character edit per 1000 characters.  The less consistent {\tt st} corpus yields a distance of 8 edits per 1000 characters. A manual inspection of Java documents formatted using  models trained with different grammars confirms that the structure of the grammar itself has little to no effect on the formatting results, at least when trained on the context features defined in \ref{features}. 

The {\tt sql} corpus shows a much higher difference between formatted files, 99 edits per 1000 characters.  Manual inspection shows that both versions are plausible, just a bit different in \NL prediction. Newlines trigger indentation, leading to bigger whitespace differences. One reason for higher edit distances could be that the noise in the less consistent SQL corpus amplifies any effect that the grammar has on formatting. More likely, the increased grammar sensitivity for SQL has to do with the fact that the {\tt sqlite} and {\tt tsql} grammars are actually for two different languages. The TSQL language has procedural extensions and is Turing complete; the {\tt tsql} grammar is 2.5x bigger than {\tt sqlite}.  In light of the different SQL dialects and noisier corpus, a larger difference between formatted SQL files is unsurprising and does not rule out grammar invariance.

\subsection{Effects of corpus size}

\begin{figure}[htbp]
\begin{center}
\scalebox{.4}{\includegraphics{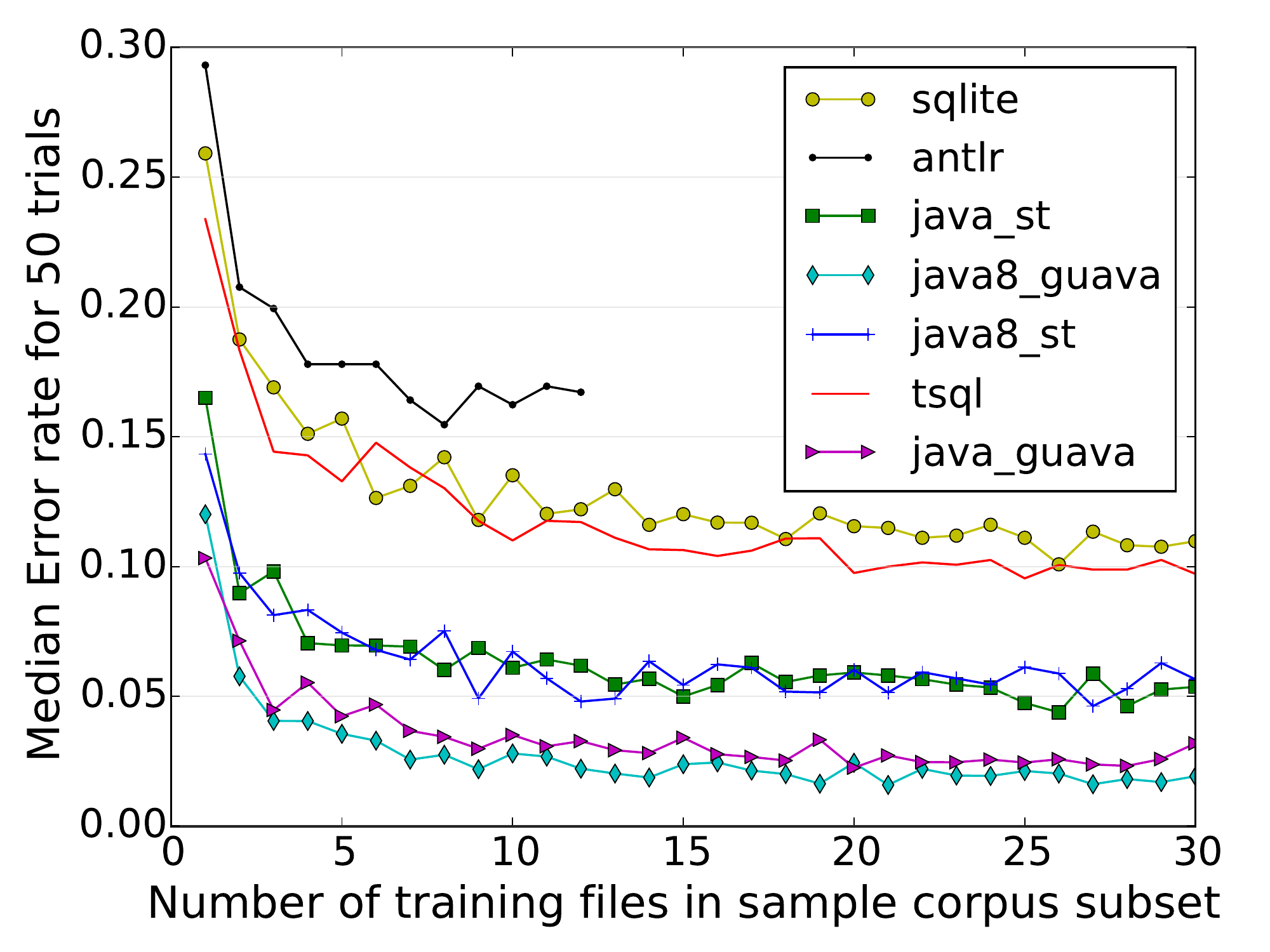}}
\vspace{-8mm}
\caption{\small Effect of corpus size on median leave-one-out validation error rate using randomly-selected corpus subsets.}
\label{corpus-size}
\end{center}
\end{figure}

Prospective users of \codebuff will ask how the size of the corpus affects formatting quality. We performed an experiment to determine: {\em (i)} how many files are needed to reach the median overall error rate and {\em (ii)} whether adding more and more files confuses the {\em kNN} classifier. Figure \ref{corpus-size} summarizes the results of an experiment comparing the median error rate for randomly-selected corpus subsets of varying sizes across different corpora and grammars. Each data point represents 50 trials at a specific corpus size. The error rate quickly drops after about 5 files and then asymptotically approaches the median error rate shown in Figure \ref{leave-one-out}. This graph suggests a minimum corpus size of about 10 files and provides evidence that adding more (consistently formatted) files neither confuses the classifier nor improves it significantly.

\subsection{Optimization and stability of model parameters}

\begin{figure}[htbp]
\begin{center}
\scalebox{.4}{\includegraphics{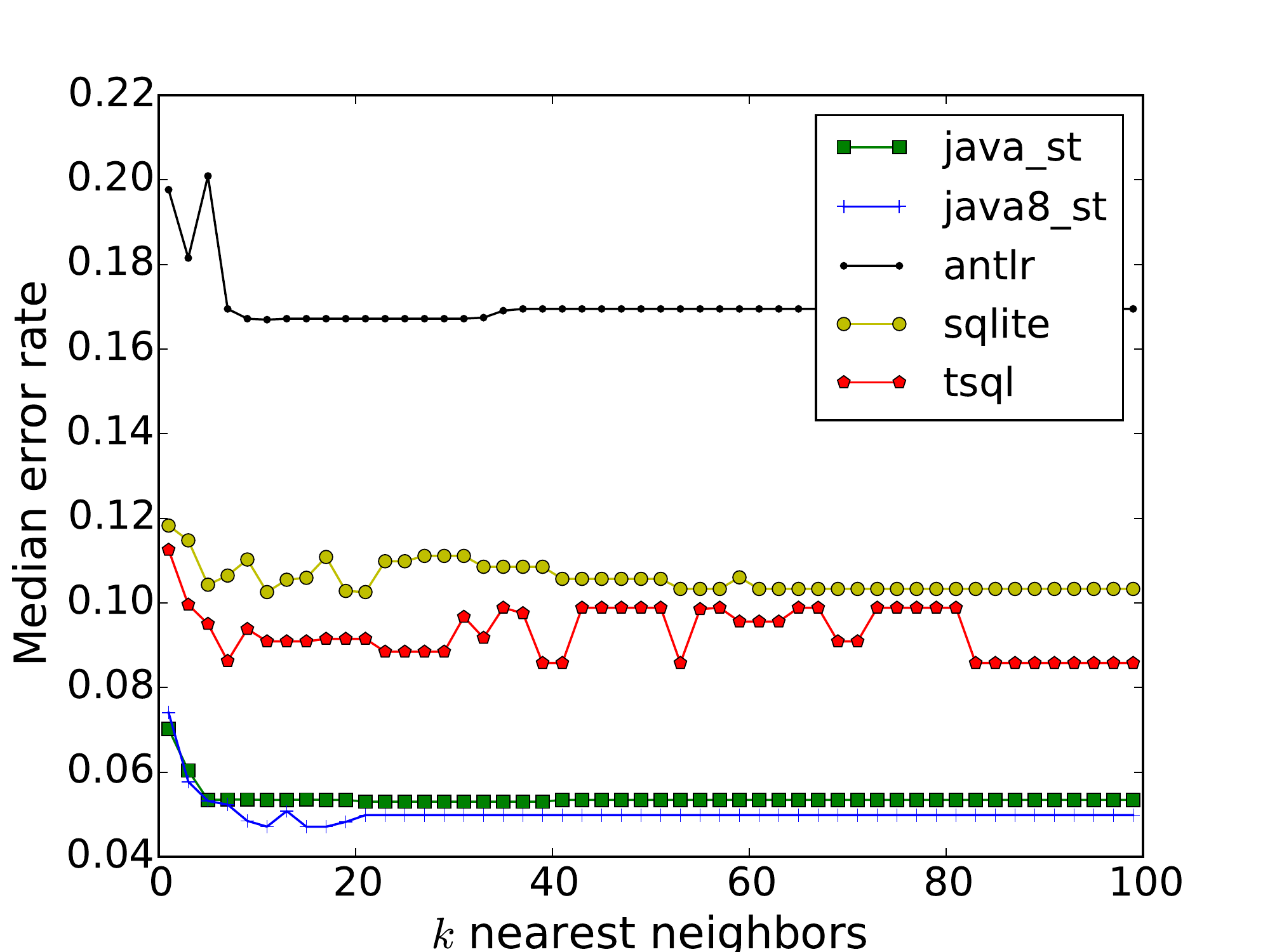}}
\vspace{-6mm}
\caption{\small Effect of $k$ on median leave-one-out error rate.}
\label{vary-k}
\end{center}
\end{figure}

Choosing a $k$ for a {\em kNN} model is more of an art but $k=\sqrt{N}$ for $N$ exemplars is commonly used. Through exhaustive manual review of formatted files, we instead arrived at a fixed $k=11$ and then verified its suitability by experiment. Figure \ref{vary-k} shows the effect of varying $k$ on the median error rate across a selection of corpora and grammars; $k$ ranged from 1 to 99. This graph supports the conclusion that a formatter can make accurate decisions by comparing the context surrounding $t_i$ to very few model exemplars.  Moreover, formatter accuracy is very stable with respect to $k$; even large changes in $k$ do not alter the error rate very much. 

\subsection{Worst-case complexity}

Collecting all exemplars to train our {\em kNN} $F_{D,G}$ model requires two passes over the input. The first pass walks the parse tree for each $d \in D$, collecting matching token pairs (Section \ref{sec:matching-tokens}) and identifying list membership (Section \ref{sec:lists}). The size of a single parse tree is bounded by the size of the grammar times the number of tokens in the document, $|G| \times |d|$ for document $d$ (the charge per token is a tree depth of at most $|P|$ productions of $G$).  To make a pass over all parse trees for the entire corpus, the time complexity is $|G| \times N$, so in $O(N)$ for $N$ total tokens in the corpus.

The second pass walks each token $t_i \in d$ for all $d \in D$, using information computed in the first pass to compute feature vectors and capture whitespace as \WS and \HPOS formatting directives.  There are $m$ features to compute for each of $N$ tokens. Most of the features require constant time, but computing the earliest ancestors and identifying tokens for indentation and alignment might walk the depth of the parse tree in the worst case for a cost of $O(\log(|G| \times |d|))$ per feature per token.  For an average document size, a feature costs $O(\log(|G| \times N/|D|))$. Computing all $m$ features and capturing whitespace for all documents costs 

\[O(m \times N \times \log(|G| \times N/|D|)) =\]
\[
 O(N \times m \times (\log(|G|) + \log(N) - \log(|D|)))
\]

\noindent which is in $O(N \log N)$. Including the first pass over all parse trees adds a factor of $N$ but does not change the overall worst-case time complexity for training. 

Formatting a document is much more expensive than training. For each token in a document, the formatter requires at least one \WS classifier function execution and possibly a second for \HPOS. Each classifier function execution requires the feature vector $X$ computation cost per the above, but the cost is dominated by the comparison of $X$ to every $X_j \in {\bf X}$ and sorting those results by distance to find the $k$ nearest neighbors.  For $n=|d|$ tokens in a file to be formatted, the overall complexity is quadratic, $O(n \times N \log N)$.  For an average document size of $n=N/|D|$, formatting a document costs $O(N^2 \log N)$. Box formatters are also usually quadratic; streaming formatters are linear (See Section \ref{sec:related}).

\subsection{Expected performance}

\codebuff is instrumented to report single-threaded CPU run-time for training and formatting time. 
We report a median $1.5 s$ training time for the (overly-large by an order magnitude) {\tt guava} corpus (511 files, 143k lines) after parsing documents using the {\tt java} grammar and a training time of $1.8 s$ with the {\tt java8} grammar.\footnote{Experiments run 20 times on an iMac17,1 OS 10.11.5, 4GHz Intel Core i7 with Java 8, heap/stack size 4G/1M; we ignore first 5 measurements to account for warmup time. Details to reproduce in github repo.} The {\tt antlr} corpus, with only 12 files, is trained within $72 ms$.

\cut{
From Speed.java runs

antlr on Java8.g4
median of [4:] training 78
median of [4:] formatting 92

java_guava on Absent.java
median of [4:] training 1603
median of [4:] formatting 15

java_guava on LocalCache.java
Loaded 511 files in 1871ms
median of [4:] training 1489
median of [4:] formatting 2780

java8_guava training of LocalCache.java
Loaded 511 files in 158019ms (2.6minutes!) java8 grammar is very slow
median of [4:] training 1875
median of [4:] formatting 2199
}

Because {\em kNN} uses the complete list of exemplars as an internal representation, the memory footprint of \codebuff is substantial. For each of $N$ tokens in the corpus, we track $m$ features and two formatting directives (each packed into a single word). The size complexity is $O(N)$ but with a nontrivial constant. For the {\tt guava} corpus, the profiler shows 2.5M $X_j \in {\bf X}$ feature vectors, consuming 275M RAM. Because \codebuff keeps all corpus text in memory for testing purposes, we observe an overall 1.7G RAM footprint.

For even a modest sized corpus, the number of tokens, $N$, is large and a naive {\em kNN} classifier function implementation is unusably slow.  Ours uses an index to narrow the nearest neighbors search space and a cache of prediction function results to make the performance acceptable. To provide worst-case formatting speed figures, \codebuff formats the biggest file ({\small LocalCache.java}) in the {\tt guava} library (22,674 tokens, 5022 lines) in median times of $2.7s$ ({\tt java}) and $2.2s$ ({\tt java8}). At \textasciitilde{}2300 lines/s even for such a large training corpus, \codebuff is usable in IDEs and other applications.  Using the more reasonably-sized {\tt antlr} corpus, formatting the 1777 line {\tt Java8.g4} takes just $70ms$ (25k lines/s). 

\section{Test of Generality\label{generalize}\label{sec:validation}}

As a test of generality, we solicited a grammar and corpus for an unfamiliar language, Quorum\footnote{\url{https://www.quorumlanguage.com}} by Andreas Stefik, and trained \codebuff on that corpus. The first look at the grammar and corpus for both the authors and the model occurred during the preparation of this manuscript. We dropped about half of the corpus files randomly to simulate a more reasonably-sized training corpus but did not modify any files.

Most files in the corpus result in plausibly formatted code, but there are a few outlier files, such as {\small HashTable.quorum}.  For example, the following function looks fine except that it is indented way too far to the right in the formatted output:

\begin{scriptsize}
\begin{verbatim}
action RemoveKey(Key key) returns Value
    ...
    repeat while node not= undefined
        if key:Equals(node:key)
            if previous not= undefined
                previous:next = node:next
            else
                array:Set(index, node:next)
            end
        ...
    end
    return undefined
end
\end{verbatim}
\end{scriptsize}

\noindent The median misclassification error rate for Quorum is very low, 4\%, on a par with the {\tt guava} model, indicating that it is consistent and that \codebuff captures the formatting style.

\section{Related work\label{sec:related}\label{sec:relatedwork}\label{sec:related_work}}

The state of the art in language-parametric formatting stems from two sources: Coutaz~\cite{box} and Oppen~\cite{oppen}. 
Coutaz introduced the ``Box'' as a basic two-dimensional layout mechanism and all derived work elaborates on specifying the mapping from syntax (parse) trees to Box expressions more effectively (in terms of meta-program size and expressiveness)~\cite{formatter-gen,tosem96,J2002b,kooiker}. 
Oppen introduced the streaming formatting algorithm in which alignment and indentation operators are injected into a token stream (coming from a syntax tree serialization or a lexer token stream). Derivative and later work~\cite{prettier,olaf,pgf} elaborates on specifying how and when to inject these tokens for different languages in different ways and in different technological spaces~\cite{so55814}. 

Oppen's streams shine for their simplicity and efficiency in time and memory, while the expressivity of Coutaz' Boxes makes it easier to specify arbitrary formatting styles. The expressivity comes at the cost of building an intermediate representation and running a specialized constraint solver.

Conceptually \codebuff derives from the Oppen school most, but in terms of expressivity and being able to capture ``naturally occurring'' and therefore hard to formalize formatting styles, \codebuff approaches the power of Coutaz' Boxes. The reason is that \codebuff matches token context much like the algorithms that map syntax trees to Box expressions. \codebuff can, therefore, seamlessly learn to \emph{specialize} for even more specific situations than a software language engineer would care to express. At the same time, it uses an algorithm that simply injects layout directives into a token stream, which is very efficient. There are limitations, however (detailed in Section~\ref{sec:design}).

Language-parametric formatters from the Box school are usually combined with \emph{default formatters}, which are statically constructed using heuristics (expert knowledge). The default formatter is expected to guess the right formatting rule in order to relieve the language engineer from having to specify a rule for every language construct (like \codebuff does for all constructs). To do this, the input grammar is analyzed for ``typical constructs'' (such as expressions and block constructs), which are typically formatted in a particular way. The usefulness of default formatters is limited though. We have observed engineers mapping languages to Box completely manually, while a default formatter was readily available. If \codebuff is not perfect, at least it generalizes the concept of a default formatter by learning from input sentences and not just the grammar of a language. This difference is the key benefit of the machine learning approach; \codebuff could act as a more intelligent default formatter in existing formatting pipelines.

PGF~\cite{pgf} by Bagge and Hasu is a rule-based system to inject pretty printing directives into a stream. The meta-grammar for these rules (which the language engineer must write) is conceptually comparable to the feature set that \codebuff learns. The details of the features are different however, and these details matter greatly to the efficacy of the learner. The interplay between the expressiveness of a meta-grammar for formatting rules and the features \codebuff uses during training is interesting to observe: we are characterizing the domain of formatting from different ends.

Since \codebuff automatically learns by example, related work that needs explicit formalizations is basically incomparable from a language engineering perspective. We can only compare the speed of the derived formatters and the quality of the formatted output. It would be possible to compare the default grammar-based formatters to see how many of their rules are useful compared to \codebuff, but {\em (i)} this is completely unfair because the default formatters do not have access to input examples and {\em (ii)} default formatters are constructed in a highly arbitrary manner so there is no lesson to learn other than their respective designer's intuitions.

Aside from the actual functionality of formatting text, related work has extended formatting with ``high fidelity'' (not losing comments)~\cite{hifi,thesis2005} and partial formatting (introducing a new source text in an already formatted context)~\cite{dejonge}. \codebuff copies comments into the formatted output but partial formatting is outside the scope of the current contribution; \codebuff could be extended with such functionality.

\section{Conclusion}
\label{sec:conclusion}

\codebuff is a step towards a universal code formatter that uses machine learning to abstract formatting rules from a representative corpus.  Current approaches require complex pattern-formatting rules written by a language expert whereas input to \codebuff is just a grammar, corpus, and indentation size.  Experiments show that training requires about 10 files and that resulting formatters are fast and highly accurate for three languages. Further, tests on a previously-unseen language and corpus show that, without modification, \codebuff generalized to a 4th language.  Formatting results are largely insensitive to language-preserving changes in the grammar and our {\em kNN} classifier is  highly stable with respect to changes in model parameter $k$.  Based on these results, we look forward to many more applications of machine learning in software language engineering. 



\bibliographystyle{abbrvnat}


\bibliography{formatting}

\end{document}